\documentclass[aps,prd,amsmath,amssymb,longbibliography,
nofootinbib,twocolumn,showpacs,preprintnumbers]{revtex4-1}
\pdfoutput=1

\usepackage[toc,page]{appendix}
\usepackage{graphics}      
\usepackage{graphicx}     
\usepackage{epsfig}
\usepackage{epstopdf}
\usepackage{caption} 
\usepackage{subcaption} 
\usepackage{float} 

\usepackage[table]{xcolor}    
\usepackage{url}           
\usepackage{bm}            
\usepackage{mathrsfs}
\usepackage{breakurl}

\usepackage{hyperref}
\usepackage{color}
\usepackage{soul}
\soulregister\cite7
\usepackage{tikz}
\usetikzlibrary{calc}

\usepackage{pgfplots}
\pgfplotsset{width=5.7cm,compat=1.9}
\pgfplotsset{yticklabel style={text width=2.5em,align=right}}
\pgfplotsset{every axis/.append style={thick}}

\newcommand\be{\begin{equation}}
\newcommand\ee{\end{equation}}
\newcommand\nono{\nonumber}
\newcommand\bse{\begin{subequations}}
\newcommand\ese{\end{subequations}}
\newcommand\bea{\begin{eqnarray}}
\newcommand\eea{\end{eqnarray}}

\newcommand{\pd}{\partial}

\newcommand\ringring[1]{%
  {
   \mathop{\kern0pt #1}\limits^{
     \vbox to-1.85ex{
       \kern-2ex 
       \hbox to 0pt{\hss\normalfont\kern.1em \r{}\kern-.45em \r{}\hss}%
       \vss 
     }
   }
  }
}
\newcommand{\q}{\quad}
\newcommand{\qq}{\qquad}

\allowdisplaybreaks
\begin{document}

\title{Encounter between an extended hyperelastic body and a Schwarzschild black hole with quadrupole-order effects}

\author{Nishita Jadoo}
\affiliation{Department of Physics, North Carolina State University, Raleigh, NC 27695 }
\altaffiliation{Present address: Department of Physics, University of Idaho, Moscow, ID 83843}
\author{J. David Brown}
\affiliation{Department of Physics, North Carolina State University, Raleigh, NC 27695}
\author{Charles R. Evans}
\affiliation{Department of Physics and Astronomy, University of North Carolina, Chapel Hill, NC 27599}
\affiliation{School of Mathematics and Statistics, University College Dublin, Belfield D04 N2E5, Dublin 4, Ireland}

\begin{abstract}
We model the general relativistic interaction of a small hyperelastic sphere with a Schwarzschild black hole 
as it follows an initially marginally-bound orbit through a close encounter.  While the interaction reveals 
effects that are encoded by the Mathisson-Papapetrou-Dixon (MPD) multipolar equations through quadrupole 
order, the calculation is made using an independent general relativistic finite element scheme that we described 
earlier (Phys.~Rev.~D 108(8):084020, October  2023).  The finite element calculation is done in Schwarzschild 
coordinates, following a large and scalable number of mass elements in interaction with each other through 
elastic forces derived from a potential energy function and with the spacetime 
geometry.  After the fact, we analyze the dynamics using a local Fermi coordinate system, computing (1) the 
deviation of the center of mass of the body relative to the initial marginally-bound orbit, (2) changes in 
orbital and spin angular momenta, and (3) the decrease in orbital energy and accompanying deposition of 
energy into internal elastic dynamics.  The interaction leads to the capture of the small body into a highly eccentric 
orbit ($e \simeq 0.99998$ in a sample calculation).
\end{abstract}

\maketitle


\section{Introduction}

The study of extended bodies in general relativity has a long history beginning with Einstein and 
Grommer \cite{EinsGrom27}.  In the test particle limit, a point particle moves along a geodesic of the given 
spacetime.  The motion of a physical, extended body deviates from a geodesic due to two effects.  First, if 
the mass cannot be ignored, the object will produce a gravitational field that alters the geometry and gives rise to a 
self-force and self-torque that affect the trajectory and spin angular  	
momentum.  Second, even in the test-mass limit, the infinitesimal pieces of the extended body move along 
separate worldlines and experience different interactions with the gravitational field.  The net result of 
these individual interactions drives changes in the center of mass motion and spin.

Subsequently, Mathisson \cite{Mathisson1937} developed equations of motion for an extended test body in curved spacetime 
using a covariant multipolar expansion.  Further work was done by Papapetrou \cite{Papapetrou1951} 
and others \cite{Pirani1956, Tulczyjew1959, Madore1969}, including the elucidation of various spin-supplementary 
conditions \cite{Costa2014}. 
The approach was extended and clarified in a series of papers by Dixon \cite{Dixon1970I, Dixon1970II, Dixon1974III}, who 
derived the laws of motion for the worldline and spin through all multipole order contributions.  The resulting equations 
are commonly referred to as the Mathisson--Papapetrou--Dixon (MPD) equations.  At zeroth order, the body moves along a 
geodesic, while the next order effect is due to the coupling of the body's spin to the spacetime curvature, producing 
both precession and deflection of the worldline.  Beyond this so-called pole-dipole approximation, the body's quadrupole 
and higher-order moments couple
to the spacetime curvature and gradients in the curvature, lending further corrections to the motion and spin dynamics.  
See reviews in \cite{Dixon1979,Dixon2015,Harte2015}.  

More recently, work has been done in using action principles to 
obtain equivalent results \cite{Bailey, Porto, Steinhoff, LB2017} and efforts have been pursued to extend the laws of motion 
to extended bodies with self-gravity \cite{RaciFlan05,Harte2012,Harte2015,Kope19,HartBlanFlan25}, where the momenta respond 
to an effective metric.  This latter approach is an extension to finite-sized objects of the familiar point-particle 
self-force problem that is widely used in first-order perturbation theory calculations of extreme-mass-ratio inspiral (EMRI)
(see reviews e.g., \cite{BaraPoun19,PounWard22}).  Numerous calculations of deviations from geodesic point-particle 
motion under combined MPD and gravitational wave emission effects during EMRI inspirals or scattering events have been 
made over the past decade
\cite{BG2017,HLBN2016,LHBN2017,WarbOsbuEvan17,MathPounWard22,LWLS2022,TLA2023,DHBH2023,MathWardPounWarb25}. 

The MPD formalism involves the choice of a worldline, with associated spatial hypersurfaces that foliate the worldtube within which the extended body is confined \cite{Dixon1974III,Harte2020}.  The integrated conservation laws constrain 
only the monopole and dipole moments.  The quadrupole and higher-order moments, which contribute to the dynamics of the linear and spin angular momenta, must be determined by added internal dynamics within the body, subject to the specifics 
of the stress-energy-momentum tensor.  See recent work \cite{Harte2020,HarteDwyer2023} on ``what is possible'' as far 
as effects of higher moments on the bulk motion.  

To close the system of equations, either the internal dynamics must be evolved, and higher moments determined dynamically, 
or a constitutive relation is imposed to place a stationary relationship between these moments and moments of the tidal field, especially the quadrupole (i.e., I-Love-Q relations; see \cite{Hinderer_2008,YY2017}) or rapid spin of the body.  
A quasistationary deformation, set by a Love-relation, will occur in the tidal field of a circular orbit binary, whereas eccentric motion or misaligned spin would generally require added dynamical evolution.

In this paper we sidestep these issues initially by making a full numerical computation of the dynamics of a small 
but extended-size elastic body in the test-mass limit.  We can therefore examine MPD effects beyond pole-dipole order 
using an encounter between a stationary (Schwarzschild) black hole and an object with full internal dynamics.  The 
dynamics is governed by a stress-energy-momentum (SEM) tensor based on a model of general relativistic hyperelasticity 
\cite{DeWitt:1962cg, Carter1972, Kijowski1992, Brown1996, Beig2003, Brown2021}. A hyperelastic model is one in which the elasticity is derived from a potential energy function, which in turn can allow large elastic deformations. We use the potential energy function for the Saint Venant-Kirchhoff model, $W(E)$, defined in Eq.~(\ref{eq:W(E)}). The numerical method was described in Ref.~\cite{JBE23}, which we based on a Lagrangian finite-element approach for 
numerically modeling extended bodies of arbitrary shape and size and for any given spacetime metric.  The scheme does not 
require the use of a tailored, local normal-neighborhood coordinate system.  

However, after the fact, as we show in this paper, it is possible to 
construct a local Fermi coordinate system based around some timelike geodesic that closely follows the extended body, 
transforming the description of the system from Schwarzschild coordinates.  It is then possible to analyze the behavior 
of the system in more familiar terms, providing estimates of spin angular momentum and energy deposition and measuring 
the deflection of the orbit and decrease in orbital energy and orbital angular momentum.  The behavior is similar to that 
found in the tidal interaction of a white dwarf star with an intermediate mass black hole \cite{PhysRevD.87.104010}, 
though in the latter calculation the dynamics is computed in the Fermi frame itself and the internal dynamics is provided 
by a balance between fluid pressure and self-gravity.

Our numerical scheme was previously validated in flat spacetime.  In this paper, we introduce general relativistic 
effects by modeling the interaction of a hyperelastic body on a marginally-bound scattering orbit (i.e., an orbit 
inbound from infinity with specific energy $\mathcal{E} = 1$) with a Schwarzschild 
black hole.  The total mass of the extended body is small enough to have negligible effect on the spacetime curvature.  
Thus there is no gravitational self-force or radiation reaction; all of the self-force derives from internal dynamics 
and (effectively) the coupling between multipole moments and the tidal field.  The timelike and rotational Killing 
vectors lead to the total mass-energy and total angular momentum 
being conserved in the test body limit. 
Additionally, we assume the object has a rest mass current, so that integrated total rest 
mass is  conserved as well.  The small body has a spherical shape in equilibrium and in the absence of any tidal field.  
In these calculations, we take it to be initially non-spinning, so that all of the angular momentum is initially in 
the orbital motion.  Expected closest approaches are set to be $r_p \simeq 9.5 - 10.5 M$, where $M$ is the mass of the 
hole.   

Because the numerical method is based on integrating a large number of coupled ordinary differential equations for the 
mass elements in the body, we are able to hold known conserved quantities constant to high precision, which can be 
calculated in either the black hole centered coordinate system or the Fermi frame.  By constructing a Fermi frame along 
a worldline moving in consort with a mass element at the center of the body, we can obtain a measure of the deflection 
of the center of mass by computing and integrating the non-geodesic acceleration of the worldline.  Beyond 
the spin-curvature force (small in this case because we take the spin to be zero initially), deflection is driven by 
the coupling between the dynamical quadrupole moment and the gradient of the quadrupole tidal field (i.e., octupole tide).  

We can similarly calculate in this frame an estimate of the spin of the body about this center of mass and 
calculate dynamic deformations of the object.  More specifically, when integrating the conserved total angular momentum 
over the object in the Fermi frame, we 
identify separate parts of the integral that represent the orbital angular momentum and the spin angular momentum, 
respectively, when the tidal interaction effects are weak (i.e., at early and late times in the encounter).  We can 
thereby follow the transfer of angular momentum into spin during the deepest parts of the encounter and a 
corresponding, balancing decrease in  the orbital angular momentum.  The coupling is primarily due to the 
anti-symmetric product between the dynamical quadrupole moment and the electric part of the Riemann tensor (i.e., the 
quadrupole tidal field).  

The same process is true with  energy.  The method is accurate enough that the difference between total energy and 
rest energy can itself be accurately followed.  The integral that computes this difference in the Fermi frame can be 
separated into parts that represent, far from the tidal interaction, the orbital kinetic and potential energy (total 
energy as if it were point-like) and internal kinetic and elastic potential energies.  We are able to follow the transfer 
of orbital energy to internal energy, which occurs principally during the deepest part of the encounter.  The effect is 
that the small body acts against the tidal field and enters a bound orbit while becoming excited into internal 
oscillations and overall 
rotation.  In the Fermi frame we find it possible to decompose the oscillations into normal modes and compare their 
oscillation frequencies to analytically-known normal mode frequencies.  The dominant mode is the circular $l=2$, $m=2$ 
bar mode that is set into rotation by the quadrupole tidal field.

Beyond the present demonstration, it is conceivable that this method of treating an extended elastic body might be 
ported to numerical relativity calculations of neutron star--black hole or neutron star--neutron star binaries, to 
model at least the early stage of inspiral.  The computational cost of such a calculation would still be dominated by 
the numerical evolution of the gravitational field but the numerical treatment of the dynamics of the material body, 
or bodies, would be efficient and accurate.  To our knowledge, something like this has not yet been attempted in 
numerical relativity.  Instead, neutron stars are primarily modeled as perfect fluids with an equation of state relating 
the pressure and energy density.  Neutron star interiors are known, however, to include a solid crust composed 
of crystallized nuclear matter.  More exotic (hybrid) neutron stars are posited to have deconfined quark matter cores, 
which may also be crystalline \cite{W1984}.  Reference \cite{LLL2017} found that including elastic quark cores in 
neutron stars can result in universal I-Love-Q relations \cite{YY2017}, relating the moment of inertia, the 
tidal deformability parameter, and the quadrupole deformation.  Early work on treating stars with a general relativistic 
model of elasticity can be found in \cite{Karlovini2003I, Karlovini2004II, Karlovini2004III, Karlovini2007IV}, with 
perturbative-level dynamics.  

White dwarf stars are also expected, depending upon age, to have crystalline interiors, with in some cases up to 
99\% of the interior in solid form \cite{Camisassa2019}.  While general relativity is a weak correction in all but 
the most massive white dwarfs, reference \cite{PC2022} included it in a study of tidal deformability of 
crystalline white dwarfs.  The present finite element model might be ported to similar such studies of tidal effects 
in white dwarfs, with either Newtonian or general relativistic gravity.

More direct follow-on work might include beginning the calculation with an initially rapidly rotating body.  Then we 
would have a direct, independent calculation of MPD spin-curvature effects.  Similarly, it might be interesting to 
extend such calculations to include a Kerr black hole, and examine even more general spin-coupling and precession 
effects.  It is also worth considering whether self-gravity in the elastic body might be included, with a perturbative 
treatment of the spacetime geometry relative to the stationary background.  In this case, the motion and spin 
dynamics would become a mix of internal elastic self-force and gravitational self-force.

The outline of this paper is as follows.  We begin in Sec.~\ref{sec:formalism} by reviewing the formalism and code we 
use for numerically modeling extended bodies in general relativity.  We consider conservation laws for a hyperelastic 
body in general spacetimes and in spacetimes with symmetry in Sec.~\ref{sec:conslaws}.  We describe how we compute 
Fermi frames about different worldlines in Sec.~\ref{sec:Fermiframes}.  We explain how we compute the deviation of 
the approximate center of mass from a timelike geodesic in Sec.~\ref{sec:localestcom} and other approximate local 
quantities in Sec.~\ref{sec:locestphy}.  We then describe how we set up initial data in Sec.~\ref{sec:ini} and 
finally describe the results of the tidal encounters in Sec.~\ref{sec:results}.  In Sec.~\ref{sec:concl}, we provide concluding remarks and further discussion of avenues for future work.

\section{Formalism and code}
\label{sec:formalism}
The formalism we use to numerically model a hyperelastic body in general relativity is described in detail in Ref.~\cite{JBE23}. Here we give a brief review. 

The general relativistic theory of elasticity was first developed in
Refs.~\cite{Carter1972,Kijowski1992,Beig2003}. We follow the Lagrangian approach as described in Ref.~\cite{Brown2021}. 
The four--dimensional spacetime manifold is denoted by $\mathcal{M}$, with spacetime coordinates $x^\mu$ and metric $g_{\mu\nu}$. The matter space $\mathcal{S}$ is the space of material points in the body. The coordinates on $\mathcal{S}$ are $\zeta^i$, where $i=1,2,3$. The worldline of the material point $\zeta^i$ is described by 
$x^\mu = X^\mu(\lambda, \zeta)$, where $\lambda$ is a worldline parameter. 
The four--velocity of the material point $\zeta^i$ is
\be 
U^\mu = \dot X^\mu/\alpha \ , 
\ee 
where the ``dot" denotes $\partial/\partial\lambda$ and
$\alpha = \sqrt{-\dot X^\mu \dot X_\mu}$ 
is the material lapse function. We also define the 
Landau--Lifshitz radar metric 
\be
    f_{ij} = X^\mu_{,i} (g_{\mu\nu} + U_\mu U_\nu) X^\nu_{,j} \ ,
\ee
where ${,i} = \partial/\partial\zeta^i$. The radar metric defines 
the proper spatial distance between points of the body as measured orthogonal to the worldlines. 

We now specialize to spacetime coordinates $t$, $x^a$ (with $a = 1,2,3$), where the $t={\rm const}$ surfaces are spacelike. We employ a 3+1 splitting to write the spacetime metric $g_{\mu\nu}$ in terms of the spatial 
metric $g_{ab}$, lapse function $N = \sqrt{-1/g^{tt}}$ and
shift vector $N_a = g_{ta}$. (Latin indices are lowered and 
raised with the spatial metric and its inverse.) 

If we choose the
worldline parameter $\lambda$ to coincide with $t$, then 
$X^0(\lambda,\zeta) = \lambda$ so that $\dot X^0 = 1$ and  $X^0_{,i} = 0$. The radar metric becomes \cite{Brown2021}
\be
     f_{ij} = X^a_{,i} (g_{ab} + \gamma^2 V_a V_b) X^b_{,j} \ ,
\ee
where 
\be
    V^a = (\dot X^a + N^a)/N \ .
\ee
The relativistic Lorentz factor is defined by $\gamma = 1/\sqrt{1 - V^a V_a}$ and the material lapse is 
$\alpha = N/\gamma$. 

With $\lambda = t$, the action  for a relativistic hyperelastic body is a functional of $X^a(\zeta,t)$ given by \cite{DeWitt:1962cg, Brown1996}
\be
    S[X] = -\int_{t'}^{t''} dt \,  \int_{\mathcal S} d^3\zeta \, 
    \sqrt{\epsilon}\alpha\rho(E) \ . \label{eq:actiontgauge}
\ee
Here, $\epsilon_{ij}$ is the ``relaxed metric" on matter space (with inverse $\epsilon^{ij}$) and $\epsilon$ is its determinant. 
The energy density, $\rho(E)$, is a function of 
\be 
E_{ij} = (f_{ij} - \epsilon_{ij})/2 \ ,
\ee
which is the Lagrangian strain tensor. Note that $\rho(E)$ is the 
energy per unit volume of the undeformed body. 

In our simulations we write the energy density as $\rho(E) = \rho_0 + W(E)$, 
where $\rho_0$ is the rest energy density 
and $W(E)$ is the potential energy density. We use the Saint Venant-Kirchhoff model 
\be
    W(E) = \frac{\lambda}{2} (\epsilon^{ij} E_{ij})^2 
    + \mu (\epsilon^{ik}\epsilon^{jl} E_{ij} E_{kl} ) \ ,\label{eq:W(E)}
\ee
where $\lambda$ and $\mu$ are the Lam\'e constants. 

The stress-energy-momentum tensor for the elastic body is given by \cite{Brown2021}
\be\label{semtensorforelastic}
 T^{\mu\nu} = (\sqrt{\epsilon}/\sqrt{f} )\left[\rho(E)  U^{ \mu}  U^{ \nu} +  S^{ij}  F^{ \mu}_i F^{ \nu}_j \right]\ , 
\ee
where $S^{ij} = \partial\rho/\partial E_{ij}$ is the second Piola--Kirchhoff stress tensor 
and $F^\mu_i = (\delta^\mu_\nu + U^\mu U_\nu) X^\nu_{,i}$ is the deformation gradient. 
Also note that $f$ is the determinant of the radar metric $f_{ij}$. 

Our numerical code solves the equations of motion obtained 
from a discrete form of the action (\ref{eq:actiontgauge}).
The action is discretized by dividing matter space into tetrahedral elements, ${\mathcal S}_E$, labeled by an index $E$ (not to be confused with the strain tensor).  For a 
material point $\zeta^i$ in 
the element ${\mathcal S}_E$, we define
\be
    X^a(t,\zeta) = \sum_{n\in \mathcal{N}(E)} X^a_{n}(t) \; \phi^E_n(\zeta)\ , \quad \zeta^i \in \mathcal{S}_E\ ,\label{eq:discX}
\ee 
where $\phi^E_n(\zeta)$ are shape functions. 
The sum is carried out over the set of nodes contained in the element $\mathcal{S}_E$. This set is denoted by $\mathcal{N}(E)$. The discrete action is
\begin{align} 
    S[X] &= \int_{t'}^{t''} dt \sum_E  \int_{\mathcal{S}_E} d^3\zeta \; \mathcal{L} \bigg(\sum_{n\in \mathcal{N}(E)} X^a_{n}(t) \; \phi_n^E(\zeta), \nono \\
    &\q\sum_{n\in \mathcal{N}(E)} \dot X^a_{n}(t) \; \phi_n^E(\zeta), \; \sum_{n\in \mathcal{N}(E)} X^a_{n}(t) \; \phi^E_{n,i} \bigg)\ , \label{eq:discaction}
\end{align}
where ${\mathcal L}(X,\dot X,X_{,i}) = -\sqrt{\epsilon}\alpha\rho(E)$
is the Lagrangian density. The action is a functional of the coordinates of each node,  $X^a_{n}(t)$.

The nodes are chosen to be located at the vertices of the tetrahedra and the shape functions are selected to be linear. A transformation is applied to convert each tetrahedron to a regular shape with one node at the origin and the other nodes displaced by one unit along the coordinate axes. Integration over a tetrahedron is obtained using a quadrature rule with quadrature points at the vertices and quadrature weights equal to $1/24$. This makes the integration of linear functions exact. 

With these choices, the integration of the Lagrangian density over each element is obtained by simply multiplying the value of ${\mathcal L}$ at each node by the quadrature weight and summing over the nodes. The contribution 
to the action containing the variable $X^a_N(t)$, for fixed node number 
$N$, is
\begin{widetext}
\be
    S_N = \frac{1}{24} \int_{t'}^{t''} dt \sum_{E\in {\mathcal R}(N)} 
    \sum_{n\in \mathcal{N}(E)} |J_{E}| {\mathcal L} \biggl( X^a_{n}(t), \; \dot X^a_{n}(t),  
    \sum_{m\in \mathcal{N}(E)} X^a_{m}(t) \; \phi^E_{m,i}  \biggr)\ .
\ee
Here, ${\mathcal R}(n)$ is the ``ring" of $n$, the set of 
elements that have $n$ as one of their nodes. Also $|J_E|$ is the 
Jacobian of the transformation of the tetrahedra to a ``regular" shape. 

The discrete action is varied with respect to $X^a_n$ and extremized to obtain the equations of motion, 
\begin{align} \label{bigMequalsFequation}
\underbrace{\sum_{E\in {\mathcal R}(n)} |J_{E}| \frac{\partial^2{\mathcal L}}{\partial \dot X^b \partial \dot X^a} \biggr|_{n,E}}_{(M_{ab})_n} \ddot X^b_n =  \underbrace{\begin{aligned}[t]
&\sum_{E\in {\mathcal R}(n)} |J_{E}| \bigg\{ 
\frac{\partial{\mathcal L}}{\partial X^a}\biggr|_{n,E}  + \sum_{m\in \mathcal{N}(E)} \frac{\partial{\mathcal L}}{\partial  X^a_{,i}}\biggr|_{m,E} \phi^E_{n,i}     
- \, \frac{\partial^2{\mathcal L}}{\partial X^b \partial\dot X^a}\biggr|_{n,E} \dot X^b_n \\
&\qq\qq\qq\qq\qq\qq\qq\qq\qq- \, \frac{\partial^2{\mathcal L}}{\partial X^b_{,i} \partial\dot X^a} \biggr|_{n,E} \sum_{m\in \mathcal{N}(E)} \dot X^b_m \phi^E_{m,i}  
\bigg\}\ , \end{aligned}}_{(F_a)_n} 
\end{align}    
where 
$|_{n,E}$ denotes evaluation at $X^a = X^a_{n}$, $\dot X^a =  \dot X^a_{n}$, and $X^a_{,i} = \sum_{m\in \mathcal{N}(E)} X^a_{m}(t) \phi^E_{m,i}$. For each value of $n$ in Eq.~(\ref{bigMequalsFequation}), the coefficient of $\ddot X^b_n$ is a $3\times 3$ matrix in the indices $a$ and $b$. 

Despite the complexity of Eq.~(\ref{bigMequalsFequation}), the set of equations can be rewritten as a system of $6N_{\mathrm{total}}$ first-order ODEs for the variables $X^a_n$ and $V^a_n = \dot X^a_n$, where $N_{\mathrm{total}}$ is the total number of nodes. The first $3N_{\mathrm{total}}$ equations are the definitions  $V^a_n = \dot X^a_n$ with $a=1,2,3$ and $n=1,\ldots,N_{\mathrm{total}}$. Denoting the coefficient of $\Ddot{X}^b_n$ in Eq.~(\ref{bigMequalsFequation}) as $(M_{ab})_n$ and the right hand side as $(F_a)_n$, the next $3N_{\mathrm{total}}$ first-order ODEs are written in matrix form as
\begin{align}    
\underbrace{\begin{bmatrix}
(M_{11})_1 & (M_{12})_1 & (M_{13})_1 & \ldots&0\\
(M_{21})_1 & (M_{22})_1 & (M_{23})_1 & \ldots&0\\
(M_{31})_1 & (M_{32})_1 & (M_{33})_1 & \ldots&0\\
\vdots\\
0          &   \ldots   &   \ldots   &  \ldots
\end{bmatrix}}_{\text{mass matrix},\, M}
\frac{d}{d t} \left(
\begin{bmatrix}
V^1_{1}\\
V^2_{1}\\
V^3_{1}\\
\vdots\\
\vdots
\end{bmatrix}\right)
=
\underbrace{\begin{bmatrix}
(F_1)_1\\
(F_2)_1\\
(F_3)_1\\
\vdots\\
\vdots\\
\end{bmatrix}}_{\text{vector},\, F}\ . \label{eq:odemat}
\end{align}
\end{widetext}
We use the following numerical techniques to solve this linear system for the time derivatives of $V^a_n$. First, we invoke MATLAB's Symbolic Math Toolbox \cite{SymbolicMathToolbox} to write the Lagrangian symbolically in terms of $X$, $\dot X$  and $X_{,i}$. We symbolically compute derivatives of the Lagrangian and convert the results to Fortran functions. These functions are used to construct the mass matrix $M$, which 
is pentadiagonal, and the vector $F$.  We then use the subroutine DGBSV from the Fortran Linear Algebra Package (LAPACK), which uses lower–upper (LU) decomposition to solve the linear system of equations (\ref{eq:odemat}) and obtain the time derivatives of $V^a_n$.

After inverting the mass matrix $M$, we use fourth-order Runge--Kutta to evolve $X^a_n$ and $V^a_n$ at discrete values of time $t$. For stability, the time-step size $\Delta t$ must satisfy the Courant condition 
$\Delta t \leq h_\mathrm{min}/C_L$, 
where $h_\textrm{min}$ is the minimum edge length of the tetrahedral elements and $C_L$ is the longitudinal sound speed of the elastic material.

\section{Conservation laws}
\label{sec:conslaws}
The rest energy for a hyperelastic body is always conserved. 
We also define conserved quantities associated with symmetries (Killing vector fields) of the spacetime. We are particularly interested in spacetimes such as Schwarzschild and Kerr that admit timelike and rotational Killing vector fields. 

\subsection{Rest energy}

The rest energy of an elastic body is conserved in curved spacetime, independent of any symmetry. Let $\rho_0$ denote the rest energy per unit (undeformed) volume of the body. If we view the body as a continuum of particles, then $\rho_0$ is the product of particle mass and number density of particles in the undeformed (relaxed) state. 

The rest energy current for an elastic body is
\be
    {\mathcal J}^\mu  = 
    \frac{\sqrt{\epsilon}}{\sqrt{f}} \rho_0 U^\mu \ ,
\ee
where   $f$ is the determinant of the radar metric $f_{ij}$. 
The factor $\sqrt{\epsilon}\rho_0/\sqrt{f}$ is the rest energy per unit physical volume of the distorted body. In Appendix \ref{appendix:conservation}  we show  that the current satisfies 
\be\label{nablaJequalszero}
\nabla_\mu {\mathcal J}^\mu = 0 
\ee
by direct calculation.  Then the conserved rest energy is
 \be
    E_{\mathrm{rest}} =   \int_{ \Sigma } d^3\Sigma_\mu 
    {\mathcal J}^\mu \ ,\label{eq:Erest0} 
\ee
where $\Sigma$ is any spacelike surface.  The volume element is 
$d^3\Sigma_\mu = -d^3\sigma \sqrt{h} n_\mu$, where $d^3\sigma$ is the coordinate 
volume on $\Sigma$, $h$ is the determinant of the induced metric on $\Sigma$, and 
$n^\mu$ is the future pointing unit normal to $\Sigma$.

\subsection{Spacetime symmetries}

Let $\xi^\mu$ denote a Killing vector field that generates a spacetime symmetry. From the Killing equation $\nabla_\mu \xi_\nu + \nabla_\nu \xi_\mu = 0$ and the local conservation  of the stress-energy-momentum tensor $T^{\nu\mu}$, we find that the current $T^{\mu\nu} \xi_\nu$ 
is conserved: $\nabla_\mu ( T^{\mu\nu} \xi_\nu) = 0$. 

Let $\xi^\mu_{(t)}$ denote the timelike Killing vector field for Schwarzschild or Kerr spacetime. The total conserved energy is 
\be
    E_{\mathrm{tot}} = -\int_\Sigma d^3\Sigma_\mu  T^{\mu\nu} \xi^{(t)}_{\nu}\ . \label{eq:Etot0}
\ee
The Killing vector field corresponding to spatial rotations in the $xy$ plane in Schwarzschild or Kerr spacetime is denoted  $\xi^\mu_{(\phi)}$. The total angular momentum in the $z$ direction, 
\be
    J_{\mathrm{tot}} =  \int_\Sigma d^3\Sigma_\mu T^{\mu\nu} \xi^{(\phi)}_{\nu} \ ,\label{eq:Jz}
\ee
is conserved.

\section{Fermi frames}
\label{sec:Fermiframes}

In this section we describe the coordinate systems used to compute local quantities for the elastic body. We refer to these as ``Fermi coordinates" or ``Fermi frames." The calculation of local quantities is carried out  after the numerical time integration described in Sec.~\ref{sec:formalism} is complete. 

The term ``Fermi normal coordinates" usually refers to a set of coordinates tied to a timelike geodesic parametrized by proper time \cite{Manasse1963}. These coordinates are inertial in the sense that the metric along the worldline 
coincides with the Minkowski metric and the Christoffel symbols vanish \cite{MTW}. The construction of Fermi normal coordinates can be generalized to an arbitrary timelike worldline parametrized by proper time \cite{Poisson:2011nh}. In this case the metric 
coincides with the Minkowski metric along the worldline, but the Christoffel symbols are 
nonvanishing---they depend on the acceleration of the worldline. 

In Appendix  \ref{FermiCoordAppendix} we review the construction of Fermi coordinates 
and collect the various relations needed for the calculations of local 
properties of the elastic body. 

For each of the Fermi coordinate systems the constant time slices are orthogonal to  a ``central worldline." 
We employ the following systems: 
\begin{enumerate}
\item Initial data frame. The central worldline for this frame is a geodesic that follows a parabolic orbit. This is used to define the initial data.  The body is placed in equilibrium under the influence of the local tidal field in the initial data frame. The data is then interpolated to the initial spacetime coordinate slice $t=0$. 
\item Fiducial frame. The central worldline is a ``fiducial node" in the discretized body. The fiducial node is the node closest to the center of mass of the body when the body is in its relaxed state. We use the Fermi frame attached to the fiducial node to compute the normal modes of oscillation of the sphere. We also use this frame to find the center of mass of the body, which is then interpolated to obtain the center of mass worldline in the spacetime coordinate system. 
\item Center of mass frame. The central worldline is the body's center of mass. We use this Fermi frame to calculate the energy and angular 
momentum of the body. 
\item Geodesic frame. The central worldline is a geodesic that initially coincides with the center of mass of the body. Beyond the initial time, the body's center of mass deviates from a geodesic. We monitor the center of mass of the body in the Fermi frame attached to this geodesic. 
\end{enumerate} 

\subsection{Fermi coordinates}\label{ssFermiCoords}

In Appendix \ref{FermiCoordAppendix} we  carry out the construction of Fermi coordinates. The central worldline,  denoted $X^\mu_{(cw)}(t)$, is not necessarily a geodesic. The 
proper time $\bar t(t)$ along the central worldline, as a function of $t$,  is obtained by integrating the 
equation $d\bar t/dt = \alpha^{(cw)}$, where 
\be
    \alpha^{(cw)} = \sqrt{-\dot X^\mu_{(cw)} \dot X_\mu^{(cw)} } 
\ee
and the dot denotes $d/dt$. We use the second--order scheme 
\be
    \bar t_{i+1} = \bar t_i + \alpha_i^{(cw)} \Delta t + \dot\alpha_i^{(cw)} \Delta t^2/2
\ee
where $i$ denotes the time step. The quantity $\dot\alpha_i^{(cw)}$ is computed using a  
4--point centered stencil (except for the first two and last two time steps, which use 
one--sided stencils). 
This calculation gives us $\bar t(t)$, the proper time as a function 
of coordinate time $t$.  

In Appendix \ref{FermiCoordAppendix} we show that  the 
spacetime coordinates $x^\mu = (t,x^a)$  and the  Fermi coordinates 
$x^{\bar \mu} = (\bar t,x^{\bar a})$  of an event ${\mathcal P}$ near the central worldline are related by 
\begin{align}\label{FermiSpTmCoordTransf}
    x^\mu(\bar t,\bar x) &= X_{(cw)}^\mu(t(\bar t))  + e^\mu_{\bar a}(t(\bar t))   x^{\bar a}
   \nono\\
   \qquad &  - \frac{1}{2} \Gamma^\mu_{\alpha\beta}(t(\bar t))  e^\alpha_{\bar a}(t(\bar t))  e^\beta_{\bar b}(t(\bar t)) x^{\bar a} x^{\bar b}  + {\mathcal O}(\bar x^3) .
\end{align}
Here, $e^\mu_{\bar a}(t(\bar t))$ (with  ${\bar a} = 1,2,3$) is an orthonormal triad of basis vectors, Fermi--Walker transported along the central worldline.  Also, $\Gamma^\mu_{\alpha\beta}(t(\bar t))$ are the Christoffel symbols evaluated on the central worldline. 

\begin{figure}
\centering
\includegraphics{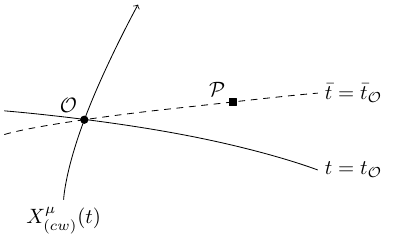}
\caption{\label{fig:timeslices1} The constant time slice $t = t_{\mathcal O}$ passes 
through the event ${\mathcal O}$ on the central 
worldline $X^\mu_{(cw)}(t)$. The dashed curve is a spacelike geodesic, orthogonal to the central worldline at ${\mathcal O}$. 
The spacelike geodesic passes through a nearby event ${\mathcal P}$.  
Events along the spacelike geodesic are assigned the Fermi coordinate time $\bar t = \bar t_{\mathcal O}$, which is the proper time at event ${\mathcal O}$ along the central worldline.}
\end{figure}

Figure~\ref{fig:timeslices1} shows the central worldline 
along with slices of constant spacetime coordinate $t$ and constant Fermi coordinate $\bar t$. 

The metric in Fermi coordinates $\bar t$, $x^{\bar a}$ is 
\begin{widetext}
\bse\label{metricinFermicoords}
\bea
     g_{\bar t \bar t} (\bar t,\bar x)  & = &    -1 - 2A_{\bar a}\bigr|_{(cw)} x^{\bar a} - \bigl[  R_{\bar t\bar a\bar t\bar b}   + A_{\bar a} A_{\bar b}  \bigr]\bigr|_{(cw)}   x^{\bar a} x^{\bar b} + {\mathcal O}(\bar x^3)    \ ,\\
     g_{\bar t \bar a} (\bar t,\bar x)   & = & \frac{2}{3}  R_{\bar t\bar b\bar c\bar a} \bigr|_{(cw)}  x^{\bar b} x^{\bar c} + {\mathcal O}(\bar x^3)  \ ,\\
     g_{\bar a \bar b} (\bar t,\bar x) & = & \delta_{\bar a \bar b} + \frac{1}{3}  R_{\bar c\bar a\bar b\bar d}\bigr|_{(cw)}  x^{\bar c} x^{\bar d} + {\mathcal O}(\bar x^3)  \ .
\eea
\ese
\end{widetext}
Here, terms such as $R_{\bar t\bar b\bar c\bar a}\bigr|_{(cw)}$ 
are the Fermi coordinate components of the Riemann tensor
evaluated along the central worldline at $x^\mu = X^\mu_{(cw)}(t(\bar t))$. Similarly, $A_{\bar a} \bigr|_{(cw)}$ are the 
Fermi coordinate components of the acceleration of the 
central worldline.

\subsection{Orthonormal triad}
Each Fermi coordinate system requires an orthonormal triad $e^\mu_{\bar a}(t)$, defined by Fermi--Walker transport along the 
central worldline. 
Let us assume that the spacetime metric is diagonal. Then  $N^a = 0$ and the metric has the form 
\be
    ds^2 = -N^2 dt^2 + \rho_1^2 (dx^1)^2 
    + \rho_2^2 (dx^2)^2 + \rho_3^2 (dx^3)^2 \ . 
\ee
We choose the tetrad legs at $t = 0$  to be
\bse\label{orthotetrad}
\begin{align}
    e^\mu_{\bar x} &=  \frac{1}{\rho_1} \delta^\mu_1
    + \frac{\rho_1}{N^2} \gamma \dot X^1_{(cw)} \eta^\mu \ ,\\
    e^\mu_{\bar y} &=  \frac{1}{\rho_2} \delta^\mu_2
    + \frac{\rho_2}{N^2} \gamma \dot X^2_{(cw)} \eta^\mu \ ,\\
    e^\mu_{\bar z} &=  \frac{1}{\rho_3} \delta^\mu_3
    + \frac{\rho_3}{N^2} \gamma \dot X^3_{(cw)} \eta^\mu \ ,
\end{align}
\ese
where  
\be
    \eta^0 = 1 \ ,\quad \eta^a = \frac{\gamma}{1 + \gamma} \dot X^a_{(cw)} \ .
\ee
The triad  is numerically evolved using the equation $U^\alpha_{(cw)} \nabla_\alpha e^\mu_{\bar a} = U^\mu_{(cw)}A_{\alpha}^{(cw)} e^\alpha_{\bar a}$ for 
Fermi-Walker transport. Here, $U^\mu_{(cw)} = \dot X^\mu_{(cw)}/\alpha_{(cw)}$ and $A^\mu_{(cw)}$ are the spacetime coordinate components of the four--velocity and acceleration of the central worldline. The Fermi-Walker transport  equation is written as 
\be
   \dot e^\mu_{\bar a} = - \dot X_{(cw)}^\alpha \Gamma_{\alpha\beta}^\mu e^\beta_{\bar a} + \dot X_{(cw)}^\mu A_\alpha^{(cw)} e^\alpha_{\bar a} \ ,
\ee
and discretized using fourth-order Runge--Kutta to give the triad components $e^\mu_{\bar a}$ at discrete  times $t$. 

\subsection{From spacetime to Fermi coordinates}

\begin{figure}
\centering
\includegraphics{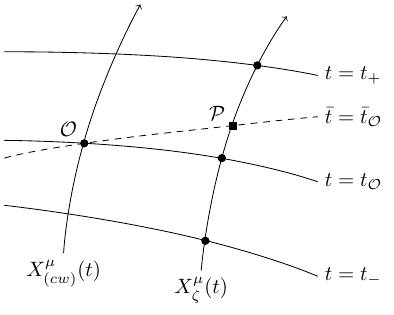}
\caption{\label{fig:timeslices}The spacelike geodesic (dashed curve) is orthogonal to the central worldline $X^\mu_{(cw)}( t)$ at  $\mathcal{O}$. The spacelike geodesic crosses the 
generic node's worldline $X^\mu_\zeta(t)$ at ${\mathcal P}$.}
\end{figure}

When working with the fiducial frame, center of mass frame, and geodesic frame, we are faced with the task of interpolating data from the spacetime coordinate grid to the Fermi frame. 
Consider the worldline $X^\mu_{\zeta}(t) \equiv X^\mu(\zeta,t)$ for some generic node in the body, 
as shown in Fig.~\ref{fig:timeslices}. This worldline is parametrized by 
coordinate time $t$, so that $X^0_\zeta(t) = t$. Because the elastic body is evolved numerically, 
we only know the worldline coordinates at discrete values of $t$.  To be precise, we know the values $X^a_{\zeta}(t_-)$, $X^a_{\zeta}(t_{\mathcal O})$ and $X^a_{\zeta}(t_+)$. These events are shown as 
filled circles in Fig.~\ref{fig:timeslices}. 

We need to compute the Fermi coordinates $x^{\bar a}$ of the event ${\mathcal P}$, which lies on the generic node worldline at the Fermi time $\bar t = \bar t_{\mathcal O}$. 
The spacetime coordinates for ${\mathcal P}$ are $(t_{\mathcal P}, X^a_\zeta(t_{\mathcal P}))$ and the Fermi coordinates for ${\mathcal P}$ are $(\bar t_{\mathcal O},x^{\bar a})$.  With this notation,  Eqs.~(\ref{FermiSpTmCoordTransf}) become 
\begin{widetext}
\bse\label{FermiSpTmCoords2}
\begin{align}
    t_{\mathcal P}  &= t_{\mathcal O} + e^0_{\bar a}(t_{\mathcal O})  x^{\bar a} - \frac{1}{2} \Gamma^0_{\alpha\beta}(t_{\mathcal O}) e^\alpha_{\bar a}(t_{\mathcal O}) e^\beta_{\bar b}(t_{\mathcal O}) x^{\bar a} x^{\bar b} + {\mathcal O}(\bar x^3)  \ ,\\
    X^a_\zeta(t_{\mathcal P}) &= X_{(cw)}^a(t_{\mathcal O}) + e^a_{\bar a}(t_{\mathcal O})  x^{\bar a}
    - \frac{1}{2} \Gamma^a_{\alpha\beta}(t_{\mathcal O}) e^\alpha_{\bar a}(t_{\mathcal O}) e^\beta_{\bar b}(t_{\mathcal O}) x^{\bar a} x^{\bar b} + {\mathcal O}(\bar x^3) \ .
\end{align}
\ese
We determine $x^{\bar a}$ as follows. First, expand the left--hand side 
of Eq.~(\ref{FermiSpTmCoords2}b) in a series about $t_{\mathcal O}$: 
\be
    X^a_\zeta(t_{\mathcal P}) = X^a_\zeta(t_{\mathcal O}) + \dot X^a_\zeta(t_{\mathcal O}) (t_{\mathcal P} - t_{\mathcal O}) 
    + \frac{1}{2} \ddot X^a_\zeta(t_{\mathcal O}) (t_{\mathcal P} - t_{\mathcal O})^2 + \cdots  \ .
\ee
\end{widetext}
Replace the $t$ derivatives with simple finite difference formulas
\bse
\begin{align}
    \dot X^a_\zeta(t_{\mathcal O}) &\approx \frac{X^a_\zeta(t_+) - X^a_\zeta(t_-)}{2\Delta t} \ , \\
    \ddot X^a_\zeta(t_{\mathcal O}) &\approx \frac{X^a_\zeta(t_+) - 2X^a_\zeta(t_{\mathcal O}) +  X^a_\zeta(t_-)}{\Delta t^2}  \ ,
\end{align}
\ese
where $\Delta t = t_+ - t_{\mathcal O} = t_{\mathcal O} - t_-$ is the time step. 
Now use Eq.~(\ref{FermiSpTmCoords2}a) to replace $(t_{\mathcal P} - t_{\mathcal O})$, and 
Eq.~(\ref{FermiSpTmCoords2}b) becomes 
\begin{widetext}
\begin{align}\label{Eqnsforxbar}
    & X_{(cw)}^a(t_{\mathcal O}) + e^a_{\bar a}(t_{\mathcal O})  x^{\bar a}
    - \frac{1}{2} \Gamma^a_{\alpha\beta}(t_{\mathcal O}) e^\alpha_{\bar a}(t_{\mathcal O}) e^\beta_{\bar b}(t_{\mathcal O}) x^{\bar a} x^{\bar b} 
    \nono \\
    &\quad = X^a_\zeta(t_{\mathcal O}) +  \frac{X^a_\zeta(t_+) - X^a_\zeta(t_-)}{2\Delta t}
    \left( e^0_{\bar a}(t_{\mathcal O})  x^{\bar a} - \frac{1}{2} \Gamma^0_{\alpha\beta}(t_{\mathcal O}) e^\alpha_{\bar a}(t_{\mathcal O}) e^\beta_{\bar b}(t_{\mathcal O}) x^{\bar a} x^{\bar b} \right) \nono \\
    &\qquad +  \frac{X^a_\zeta(t_+) - 2X^a_\zeta(t_{\mathcal O}) +  X^a_\zeta(t_-)}{2\Delta t^2} 
    \left(e^0_{\bar a}(t_{\mathcal O})  x^{\bar a}   \right)^2  \ .
\end{align}
\end{widetext}
Note that the terms of order ${\mathcal O}(\bar x^3)$ and higher have been dropped. 
Equations~(\ref{Eqnsforxbar})  are solved with a nonlinear solver for $x^{\bar a}$. If $t_{\mathcal O}$ corresponds to the first time step, we use the values at $t_{\mathcal O}$, $t_+$ and at the third time step $t_{++}$. The central finite difference scheme 
is replaced by a forward finite difference. 

\subsection{Fermi velocity}

The local quantities that describe the elastic body depend on the four--velocity of the 
generic node at Fermi time $\bar t = \bar t_{\mathcal O}$. 
We seek the vector $\dot X^\mu_\zeta(t)$ at point ${\mathcal P}$ in Fig.~\ref{fig:timeslices}. 

Recall that the coordinate velocity $\dot X^a_\zeta(t)$ of each node is obtained from the main time evolution 
described in Sec.~\ref{sec:formalism}. Thus, we have values for $\dot X^a_\zeta(t_-)$, $\dot X^a_\zeta(t_{\mathcal O})$, 
and $\dot X^a_\zeta(t_+)$. The coordinate velocity  at point ${\mathcal P}$ is 
approximated by 
\begin{widetext}
\be 
\dot X^a_\zeta(t_{\mathcal P}) \approx \dot X^a_\zeta(t_{\mathcal O}) + \frac{ \dot X^a_\zeta(t_+) - \dot X^a_\zeta(t_-)}{2 \Delta t} (t_{\mathcal P}-t_{\mathcal O}) + \frac{1}{2} \frac{\dot X^a_\zeta(t_+)- 2\dot X^a_\zeta(t_{\mathcal O}) + \dot X^a_\zeta(t_-)} {\Delta t^2} (t_{\mathcal P}-t_{\mathcal O})^2  \ ,
\ee
\end{widetext}
with the value of $(t_{\mathcal P} - t_{\mathcal O})$ obtained from Eq.~(\ref{FermiSpTmCoords2}a). 

The four--velocity of the generic node at point ${\mathcal P}$, in spacetime coordinates, is found by normalization. That is, $U^\mu = \dot X^\mu_\zeta /\alpha_\zeta$
where $\alpha_\zeta = \sqrt{-\dot X^\mu_\zeta \dot X^\nu_\zeta g_{\mu\nu}}$. Since
the worldline parameter is coordinate time $t$, we have $\dot X^t_\zeta(t) = 1$ and the 
four--velocity reduces to 
\be
    U^t = 1/\alpha \ ,\quad U^a = \dot X^a_\zeta /\alpha  \ ,
\ee
where $\alpha = N\sqrt{1 - V^a V_a}$ with $V^a = (\dot X^a_\zeta + N^a)/N$. 
The Fermi--coordinate components of the four--velocity at ${\mathcal P}$ are obtained from the transformation law 
\bse
\begin{align}
    U^{\bar t} &= \frac{\partial\bar t}{\partial x^\mu} U^\mu \ ,\\
    U^{\bar a} &= \frac{\partial x^{\bar a}}{\partial x^\mu} U^\mu \ ,
\end{align}
\ese
using Eqs.~(\ref{inversecoordtransformation}).

\section{Local estimate of center of mass}
\label{sec:localestcom}
\subsection{Fiducial Fermi frame}\label{sec:FFF}
The center of mass worldline for the elastic body can be computed using the Fermi frame attached to the fiducial worldline. We assume that the spacetime is approximately flat in the small region corresponding to the world tube of the body. This  allows us to 
use the special relativistic relations for momentum, angular momentum, and center of  mass of an extended body. 

We begin by defining the total momentum $P^\mu$ and the angular momentum $J^{\mu\nu}(x_0)$
about a spacetime event $x_0^\mu$: 
\bse\label{flatsptmdefinitionsforPandJ}
\begin{align}
P^\mu &= \int_\Sigma d^3\Sigma_\mu  T^{\mu\nu} \label{eq:Pmu} \ ,\\ 
J^{\mu\nu}(x_0) &= 2 \int_\Sigma d^3\Sigma_\alpha ( x^{[\mu} -  x_0^{[\mu} ) T^{\nu]\alpha}\ , \label{eq:Jmunu}
\end{align}
\ese
These formulas assume flat spacetime with Minkowski coordinates. 
Here $T^{\mu\nu}$ is the stress--energy--momentum (SEM) tensor of the body. The integrations 
in Eqs.~(\ref{flatsptmdefinitionsforPandJ}) can be carried out over any spacelike hypersurface $\Sigma$. Using Stokes' theorem, 
one can show that $P^\mu$ and $J^{\mu\nu}(x_0)$ are independent of $\Sigma$. 

\begin{figure}
\centering
\includegraphics{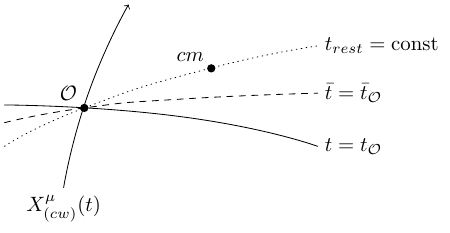}
\caption{\label{fig:timeslices3}The central worldline $X^\mu_{(cw)}(t)$ is attached to the fiducial node. The 
momentum and angular momentum of the body are computed on the slice of constant 
Fermi time $\bar t = \bar t_{\mathcal O}$. These quantities are used to compute the center of mass event  which lies on the  slice $t_{rest} = {\rm const}$. }
\end{figure}

Figure~\ref{fig:timeslices3} shows the central worldline $X^\mu_{(cw)}( t)$ (attached to the fiducial node) and a constant spacetime 
coordinate surface $t = t_{\mathcal O}$. The surface of constant Fermi time is labeled $\bar t = \bar t_{\mathcal O}$. 
We construct the Fermi coordinates and velocities of each node in the body, as described in 
Sec.~\ref{sec:Fermiframes}. Using this data we compute the Fermi--coordinate components of $P^\mu$ and $J^{\mu\nu}(x_0)$, 
where $\Sigma$ is the Fermi time slice  $\bar t = \bar t_{\mathcal O}$.  
The point $x_0^\mu$ is chosen at event ${\mathcal O}$, which is the spatial origin
of the Fermi coordinate system ($\bar t = \bar t_{\mathcal O}$ and $x^{\bar a} = 0$). 

The ``rest frame" of the body is defined by spatial hypersurfaces orthogonal to the 
momentum $P^\mu$. That is, the unit normal $n^\mu$ of a rest slice is proportional 
to $P^\mu$. The proportionality constant is the mass, $P^\mu = M n^\mu$, so that  
$P^\mu P_\mu = -M^2$. In Fig.~\ref{fig:timeslices3},  the rest slice labeled $t_{rest} = {\rm const}$  passes through the event ${\mathcal O}$. 

The center of mass on the slice $t_{rest} = {\rm const}$ is  defined by 
\be\label{definecm}
    x_{cm}^\mu = -\frac{1}{M} \int_{rest} d^3\Sigma_\alpha T^{\alpha\beta} n_\beta x^\mu  \ .
\ee
It is related to $P^\mu$ and $J^{\mu\nu}(x_0)$ 
as follows.  First write the angular 
momentum as   
\begin{align}
    J^{\mu\nu}(x_0) &= 2 \int_{\Sigma} d^3\Sigma_\alpha ( x^{[\mu} -  x_{cm}^{[\mu} ) T^{\nu]\alpha}  \nono\\
   & +  2 \int_{\Sigma} d^3\Sigma_\alpha ( x^{[\mu}_{cm} -  x_0^{[\mu} ) T^{\nu]\alpha}
\end{align}
The first term is the angular momentum about the center of mass,  $J^{\mu\nu}(x_{cm})$.
We will denote this by $S^{\mu\nu}$, for spin. The second term simplifies to  
$2 ( x^{[\mu}_{cm} -  x_0^{[\mu} ) P^{\nu]}$. 
Now contract this equation with $P_\nu$ to obtain
\be\label{JPintermediateeqn}
    J^{\mu\nu}(x_0) P_\nu = S^{\mu\nu}P_\nu + (x_{cm}^\mu - a^\mu)P^\nu P_\nu 
    - (x^\nu_{cm} - x_0^\nu)P_\nu P^\mu \ .
\ee
The last term vanishes because $x^\nu_{cm} - x_0^\nu$ (which lies in the 
$t_{rest} = {\rm const}$ slice) is orthogonal to $P_\nu$ (which is 
normal to the slice.) 

The first term, $S^{\mu\nu} P_\nu$, vanishes as well. We can see this by writing 
the momentum as $P^\mu = M n^\mu$ and evaluating the integral for spin on the rest slice. 
This gives 
\begin{align}
    S^{\mu\nu} P_\nu = M\int_{rest} d^3\Sigma_\alpha \biggl[& (x^\mu - x^\mu_{cm} ) T^{\nu\alpha} n_\nu  \nono\\
    & - (x^\nu - x^\nu_{cm} ) T^{\mu\alpha} n_\nu \biggr]  \ .
\end{align}
The second term vanishes because (on the rest slice) $x^\nu - x^\nu_{cm}$ is 
orthogonal to $n_\nu$. This leaves 
\begin{align}\label{SPintermediateeqn}
    S^{\mu\nu} P_\nu = & M\int_{rest} d^3\Sigma_\alpha x^\mu T^{\nu\alpha} n_\nu \nono\\
    & - M x^\mu_{cm} \int_{rest}  d^3\Sigma_\alpha T^{\nu\alpha} n_\nu   \ .
\end{align}
By the definition (\ref{definecm}) above,   the first term is  $-M^2 x^\mu_{cm}$. 
From the definition (\ref{flatsptmdefinitionsforPandJ}a) for momentum, we see that the  
second term simplifies to $-Mx^\mu_{cm} P^\nu n_\nu = M^2 x^\mu_{cm}$. Thus, the 
two terms in Eq.~(\ref{SPintermediateeqn}) cancel. This shows that 
the momentum and spin are orthogonal,  $S^{\mu\nu}P_\nu = 0$. 

Returning to Eq.~(\ref{JPintermediateeqn}) and solving for the center of mass, we find
\be\label{eqn:cmequation}
    x^\mu_{cm} = x_0^\mu - \frac{1}{M^2} J^{\mu\nu}(x_0) P_\nu \ .
\ee
This result allows us to determine the coordinates of the center of mass event that lies on the slice $t_{rest} = {\rm const}$ by evaluating $P_\mu$ and $J^{\mu\nu}(x_0)$ with the data on the Fermi time slice $\bar t = \bar t_{\mathcal O}$. 

We first compute the Fermi coordinate components $T^{\bar\mu\bar\nu}$ of the stress--energy--momentum tensor using the Fermi coordinates and velocities of the nodes. (Here, $\bar \mu$ ranges over the Fermi 
coordinate indices $\bar t$ and $\bar a$.) 
With the approximation that the spacetime is flat, the  volume element in Fermi coordinates reduces to $d^3\Sigma_{\bar\alpha} = d^3\bar x \, \delta_{\bar \alpha}^{\bar t}$ 
on the slice $\bar t = \bar t_{\mathcal O}$. Then the integrals for momentum and angular momentum 
become 
\bse 
\begin{align}
P^{\bar\mu} &=  \int_{\bar t = \bar t_{\mathcal O}} d^3\bar x \,    T^{\bar t\bar\mu} \ ,\\ 
J^{\bar\mu\bar\nu}(x_{\mathcal O}) &= 2 \int_{\bar t = \bar t_{\mathcal O}} d^3\bar x \,   ( x^{[\bar \mu} - x_{\mathcal O}^{[\bar \mu} ) T^{\bar\nu]\bar t} \ .
\end{align}
\ese 
Here, we have chosen $x^\mu_0 = x^\mu_{\mathcal O}$. 
Because $x^{\bar a}_{\mathcal O} = 0$  and the integral is taken over the slice $\bar t = \bar t_{\mathcal O}$, the factor $x^{\bar \mu} - x^{\bar\mu}_{\mathcal O}$ in the angular momentum has components $\bar t - \bar t_{\mathcal O} = 0$ and $x^{\bar a} - x^{\bar a}_{\mathcal O} = x^{\bar a}$. 

Finally, we convert these results to integrals over the matter space, 
\bse 
\begin{align}
P^{\bar\mu} &=  \int_{\mathcal{S}} d^3\zeta \, \mathrm{det}(x^{\bar a}_{,i})  \, T^{\bar t\bar\mu} \label{eq:totalp} \ ,\\ 
J^{\bar\mu\bar\nu}(x_{\mathcal O}) &= 2 \int_{\mathcal{S}} d^3\zeta\, \mathrm{det}(x^{\bar a}_{,i}) \,  ( x^{[\bar \mu} - x_{\mathcal O}^{[\bar \mu} ) T^{\bar\nu]\bar t} \ ,
\end{align}
\ese 
where $\mathrm{det}(x^{\bar a}_{,i})$ is the determinant of the Jacobian of the transformation. 
The factor $\mathrm{det}(x^{\bar a}_{,i})$ is computed numerically by expanding the Fermi coordinates in 
element ${\mathcal S}_E$ in terms of the shape functions, 
\be\label{xabarexpanded}
    x^{\bar a}(\bar t,\zeta) = \sum_{n\in \mathcal{N}(E)} x^{\bar a}_{n}(\bar t) \; \phi^E_n(\zeta)\ , \quad \zeta^i \in \mathcal{S}_E\ ,
\ee 
analogous to Eq.~(\ref{eq:discX}). The 
quantities $x^{\bar a}_n$ are the Fermi coordinates of the 
nodes of element ${\mathcal S}_E$. Differentiating 
Eq.~(\ref{xabarexpanded}) yields
\be
    x^{\bar a}_{,i} = \sum_{n\in \mathcal{N}(E)} x^{\bar a}_{n}(\bar t) \; \frac{\partial\phi^E_n(\zeta)}{\partial\zeta^i} \ .
\ee 
This result gives $x^{\bar a}_{,i}$ in terms of the Fermi coordinates $x^{\bar a}_n(\bar t)$ 
of the nodes. 

With the results $P^{\bar\mu}$ and $J^{\bar\mu\bar\nu}(x_{\mathcal O})$ for the momentum and 
angular momentum, we can evaluate Eq.~(\ref{eqn:cmequation}) in 
Fermi coordinates to obtain the center of mass in Fermi coordinates, $x^{\bar\mu}_{\mathrm{cm}}$.

\subsection{Deviation of the center of mass from a geodesic} \label{Roadmap}

The worldline of the center of mass is expected to be very close to the geodesic that starts out with the same coordinates and velocity as the center of mass. To quantify this small deviation, it is reasonable to compute the center of mass in the Fermi frame carried by the geodesic.

The following describes the road map we follow to obtain the deviation of the center of mass from a geodesic.
\begin{itemize}
\item Compute the coordinates of the center of mass in the fiducial Fermi frame as described in the previous subsection. This gives  $ x^{\bar\mu}_{\mathrm{cm}}$ on a set of 
discrete $t_{rest}={\rm const}$ slices. These slices cross the central worldline 
at the discrete times $t={\rm const}$ used in the numerical evolution. 

\item Interpolate the values of $x^{\bar a}_{cm}$ at $t_{rest} = {\rm const}$ to obtain $ x^{\bar a}_{cm}$ at the discrete values of $\bar t = {\rm const}$ using sixth order Lagrange interpolation. 

\item Use Eq.~(\ref{FermiSpTmCoordTransf}) to convert $ x^{\bar\mu}_{\mathrm{cm}}$ at the discrete values of $\bar t$ to spacetime coordinates.

\item Evolve a geodesic with initial data equal to the initial spacetime coordinates and velocities of the center of mass.

\item Evolve a triad along this geodesic using parallel transport and 
define a geodesic Fermi frame. 
    
\item Compute the Fermi coordinates of the center of mass in the geodesic Fermi frame. The 
calculation follows the analysis of the previous subsection, but with the fiducial Fermi frame replaced by the geodesic Fermi frame. 
\end{itemize}

The calculation of the center of mass is most accurate when the body is far from the black hole, 
where the spacetime is close to flat. We choose a scattering orbit, with 
the starting and ending points of the simulation far from the black hole. This allows us to 
obtain  accurate values for the final deviation of the center of mass from a geodesic. 

\section{Local estimates of physical quantities }
\label{sec:locestphy}
The conserved quantities described in Sec.~\ref{sec:conslaws} can be computed using the spacetime coordinates and velocities of the nodes. These quantities can also be computed in a Fermi frame. We use the Fermi frame carried by the center of mass because this leads to a natural split of the angular momentum into  orbital and spin parts. 

We carry out the computations to first order in $A^\mu$ and second order in $\bar x^a$. In particular, 
we approximate the metric components in the center of mass Fermi frame by Eqs.~(\ref{metricinFermicoords}), but drop the terms quadratic in the acceleration. As we will see, the acceleration of the center of mass is relatively small, so the quadratic terms can be safely ignored. 

The normal $n_{\bar \mu}$ to the slices of constant center of mass Fermi time are proportional to $\delta_{\bar\mu}^{\bar t}$. The normalization factor is $-(-g^{\bar t\bar t})^{-1/2}$, with the Fermi components of the inverse metric listed in Eqs.~(\ref{invmetricinFermicoords}). 
This yields
\begin{widetext}
\be
    n_{\bar\mu}(\bar t,\bar x) = -\left[ 1 + A_{\bar a}\bigr|_{(cw)}  x^{\bar a} + 
    \frac{1}{2}  R_{\bar t\bar a\bar t\bar b}\bigr|_{(cw)} \bar x^a \bar x^b \right] \delta_{\bar\mu}^{\bar t} 
    + {\mathcal O}(A^2,\bar x^3) 
\ee
\end{widetext}
to first order in $A^\mu$ and second order in $x^{\bar a}$. 

\subsection{Orbital and internal energy}
The rest energy and total energy are defined in Eqs.~(\ref{eq:Erest0}) and (\ref{eq:Etot0}). 
These quantities are computed in the center of mass Fermi frame: 
\bse 
\begin{align}
E_{\mathrm{rest}} &=  - \int_{\bar t = \bar t_{\mathcal O}} d^3\bar x \sqrt{\bar g} \, (\sqrt{\epsilon}\rho_0/\sqrt{f})  U^{\bar \mu}  n_{\bar \mu} \label{eq:Erest} \ , \\
E_{\mathrm{tot}} &= \int_{\bar t = \bar t_{\mathcal O}} d^3\bar x \sqrt{\bar g} \,  n_{\bar \mu} T^{\bar\mu\bar\nu}  \xi^{(t)}_{\bar \nu} \label{eq:Etot} \ ,
\end{align}
\ese 
Here, $\bar g \equiv {\rm det}(g_{\bar a\bar b})$ is the determinant of the 
{\em spatial} metric in Fermi coordinates: 
\be
    \sqrt{\bar g} = 1 - \frac{1}{6} R^{\bar a}{}_{\bar c\bar a\bar d} x^{\bar c} x^{\bar d} + {\mathcal O}(\bar x^3) \ .
\ee
The covariant components of the Killing vector field in Fermi coordinates, $\xi_{\bar \mu}^{(t)}$, are found from the spacetime components $\xi_\mu^{(t)}$
using Eqs.~(\ref{covvecinFermicoords}) in Appendix \ref{FermiCoordAppendix}.

We subtract the rest energy from the total conserved energy to obtain
\begin{widetext}
\be
    E_{\mathrm{tot}}-E_{\mathrm{rest}} = \int_{\bar t = \bar t_{\mathcal O}} d^3\bar x \sqrt{\bar g} \,  n_{\bar\mu} \left[ T^{\bar\mu\bar\nu} \xi^{(t)}_{\bar\nu} 
    + (\sqrt{\epsilon}\rho_0/\sqrt{f}) U^{\bar\mu} \right] \ ,
\ee
which is also conserved. 
Next, we insert the SEM tensor of the elastic body from Eq.~(\ref{semtensorforelastic}) 
and separate the result into three parts,
\begin{align}
    E_{\mathrm{tot}}-E_{\mathrm{rest}} &= \underbrace{\int_{\bar t = \bar t_{\mathcal O}} d^3\bar x \sqrt{\bar g} \, (\sqrt{\epsilon}\rho_0/\sqrt{f}) n_{\bar\mu}  U^{\bar\mu}  \left[ U^{\bar\nu}\xi^{(t)}_{\bar\nu}  +  1 \right]}_{(T + U)_\mathrm{orb} + T_\mathrm{int}}  \underbrace{+\int_{\bar t = \bar t_{\mathcal O}} d^3\bar x \sqrt{\bar g} \,( \sqrt{\epsilon}W/\sqrt{f})  n_{\bar\mu} U^{\bar \mu}  U^{\bar\nu}\xi^{(t)}_{\bar\nu}}_{U_{\mathrm{int}}} \nono\\
    & \underbrace{- \int_{\bar t = \bar t_{\mathcal O}} d^3\bar x \sqrt{\bar g} \,(\sqrt{\epsilon} S^{ij}/\sqrt{f})  n_{\bar\mu}  F^{\bar\mu}_i  F^{\bar\nu}_j \xi^{(t)}_{\bar\nu}}_{E_{\mathrm{rel}}}.  \label{eq:E3parts}
\end{align}
\end{widetext}
The first part labeled $(T + U)_\mathrm{orb} + T_\mathrm{int}$ can be identified as the sum of the orbital kinetic and potential energy and the internal kinetic energy  of the elastic body. The second part labeled $U_{\mathrm{int}}$ is the internal potential energy of the elastic body. The third part labeled  $E_{\mathrm{rel}}$ is a relativistic effect and is the contribution of spatial stress to the energy. 

We can separate $(T + U)_\mathrm{orb} + T_\mathrm{int}$ into orbital energy and internal kinetic energy parts by defining the orbital energy $(T+U)_{\mathrm{orb}}$. Here, we consider three ways of defining the orbital energy.

Firstly, we define the orbital energy as the sum of the orbital energy of the deformed elements of the elastic body. Approximate each element as a point particle moving on its own worldline,  with rest energy equal to the rest energy of the deformed element. A point particle of rest energy $m_0$ and four-velocity $U^\mu$ has ``energy-at-infinity'' equal to $-m_0 U^\mu \xi^{(t)}_\mu$ \cite{MTW}. We define the total ``energy-at-infinity'' of the elements  in the center of mass Fermi frame, 
\be 
 (T + U)_{\mathrm{orb}, \,\mathrm{sum}} = -\int d^3 \bar x \, \sqrt{\bar g} (\sqrt{\epsilon} \rho_0/\sqrt{f} )  U^{\bar \mu}   \xi^{(t)}_{\bar \mu} - E_{\mathrm{rest}}\label{eq:Eorb1} \ ,
\ee
with the rest energy $E_{\rm rest}$ removed. 
To compute $(T + U)_{\mathrm{orb}, \,\mathrm{sum}}$,  we convert it to an integral over matter space as described 
in subsection \ref{sec:FFF}. The four--velocity of each element is found from the shape functions, as in 
Sec.~\ref{sec:formalism}. 

Secondly, we define the orbital energy as
\be
(T+ U)_{\mathrm{orb}, \, \mathrm{cm}}  = E_{\mathrm{rest}}( -U^\mu_{(cw)} \xi^{(t)}_\mu - 1)\label{eq:Eorb2} \ .
\ee
This is the energy of a single particle of mass $E_{\mathrm{rest}}$ moving along the center of mass worldline, with $E_{\rm rest}$ removed. 

Lastly, we define the orbital energy by assuming that the metric in the center of mass Fermi frame is approximately flat inside the world tube of the body. 
The momentum of each element is found by assuming the element is a point particle of mass equal to its rest energy. The total momenta are then the sum of the momenta of each element. The energy is found by computing the scalar product of the total momentum and the timelike Killing vector field evaluated at the center of mass worldline. With the rest energy removed, 
\be 
(T+ U)_{\mathrm{orb}, \, P}  = - \xi^{(t)}_{\bar \mu} \bigr|_{(cw)} \int d^3 \bar x \, \sqrt{\bar g} (\sqrt{\epsilon}\rho_0/\sqrt{f} )  U^{\bar \mu}   - E_{\mathrm{rest}}\label{eq:Eorb3} \ .
\ee 
As with $(T + U)_{\mathrm{orb}, \,\mathrm{sum}}$, 
we compute this energy as an integral over matter space.

\subsection{Orbital and spin angular momentum}

We compute the conserved total angular momentum in the $z$--direction using a $\bar t =\mathrm{const}$ slice in the center of mass Fermi frame,
\be
J_{\rm tot} = -\int_{\bar t = \bar t_{\mathcal O}} d^3\bar x  \sqrt{\bar g} \,  n_{\bar\mu}  T^{\bar\mu\bar\nu} \xi^{(\phi)}_{\bar\nu} \label{eq:Jz} \ .
\ee
The Killing vector field corresponding to spatial rotations in Schwarzschild isotropic coordinates is $\xi^\mu_{(\phi)} = (0,-y,x,0)$. After lowering the index, we find the Fermi components of the covector field $\xi_\mu^{(\phi)}$ from Eqs.~(\ref{covvecinFermicoords}). The result is simplified using the identity $\nabla_\mu \nabla_\nu \xi_\alpha = R^\beta{}_{\mu\nu\alpha} \xi_\beta$, which holds for any Killing vector 
field $\xi^\mu$. 
We then insert the Fermi components of the Killing vector and unit normal covectors into Eq.~(\ref{eq:Jz}) and separate $J_{\rm tot}$ into the following parts,
\begin{widetext}    
\begin{align}
J_{\rm tot} &= 
\underbrace{ \xi^{(\phi)}_{\bar t} \bigr|_{(cw)} \int d^3 \bar x  \sqrt{\bar g} \, T^{\bar t\bar t}}_{J_1 \equiv J_{\mathrm{orbit}}}    
+\underbrace{ \xi^{(\phi)}_{\bar a}\bigr|_{(cw)} \int d^3\bar x   \sqrt{\bar g} \, T^{\bar t\bar a}}_{J_2} 
+ \underbrace{ \left( \nabla_{\bar a} \xi^{(\phi)}_{\bar t} + 2 A_{\bar a} \xi^{(\phi)}_{\bar t} \right)\Bigr|_{(cw)} \int d^3 \bar x  \sqrt{\bar g} \, x^{\bar a}  T^{\bar t\bar t}}_{J_3}  \nono\\
&+ \underbrace{\nabla_{\bar a} \xi^{(\phi)}_{\bar b} \bigr|_{(cw)}  \int d^3 \bar x  \sqrt{\bar g} \,  x^{\bar a} T^{\bar b\bar t}}_{J_4 \equiv J_{\mathrm{spin}}} 
+ \underbrace{ A_{\bar a} \xi^{(\phi)}_{\bar b} \bigr|_{(cw)} 
\int d^3 \bar x  \sqrt{\bar g} \,  x^{\bar a} T^{\bar b\bar t} }_{J_5}
\nono \\
&+ \underbrace{ \frac{1}{2}\left( 
3R^{\bar t}{}_{\bar a \bar b \bar t}\xi^{(\phi)}_{\bar t}  +  2R^{\bar c}{}_{\bar a \bar b \bar t}\xi^{(\phi)}_{\bar c}  + 4 A_{\bar a} \nabla_{\bar b} \xi^{(\phi)}_{\bar t} \right)\Bigr|_{(cw)}\int d^3 \bar x  \bar \; \sqrt{\bar g} \, x^{\bar a} x^{\bar b} T^{\bar t\bar t}}_{J_6} \nono \\
&+ \underbrace{\frac{1}{6} \left(
4R^{\bar t}{}_{\bar b \bar c \bar a} \xi^{(\phi)}_{\bar t} +  4R^{\bar d}{}_{\bar b \bar c \bar a} \xi^{(\phi)}_{\bar d} + 3\bar R_{\bar t\bar b\bar t\bar c} \xi^{(\phi)}_{\bar a} + 6 A_{\bar b} \nabla_{\bar c} \xi^{(\phi)}_{\bar a} \right)\Bigr|_{(cw)} \int d^3 \bar x   \sqrt{\bar g} \,  x^{\bar b} x^{\bar c}  T^{\bar t\bar a}}_{J_7} 
+{\mathcal O}(A^2,\bar x^3) \ . \label{eq:Qrotparts}
\end{align}
Each integral is taken over a constant $\bar t$ slice.
\end{widetext}

Let us consider each term in $J_{tot}$ separately. 
\begin{itemize}
\item[$J_1$:] We identify this term as the orbital angular momentum. Note that the 
integral in $J_1$ is  the total mass of the elastic body (call it ${\mathcal M}$) as 
seen in the inertial frame. Also, $\xi^{(\phi)}_{\bar t} = \xi_\mu^{(\phi)} U^\mu$  where $U^\mu$ is the 
four--velocity of the center of mass. Since the center of mass worldline is 
approximately geodesic, the inner product $\xi_\mu^{(\phi)} U^\mu$
is approximately the specific orbital angular momentum ${\mathcal L}$ of a 
point particle moving along the center of mass worldline. Therefore $J_1$ 
is the product ${\mathcal M}{\mathcal L}$. For our simulations ${\mathcal M} \sim 10^{-18}M$ and 
${\mathcal L} \sim 10 M$, which yields $J_1 \sim 10^{-17} M^2$. 
\item[$J_2$:] This term is very small because the integral is 
very small---it is the momentum of the elastic body in the center of mass frame. The center of mass condition is $\int d^3\bar x\, T^{\bar t\bar t} x^{\bar a} = 0$ in flat spacetime. Differentiate this expression with respect to $\bar t$, use the 
conservation law $\partial_{\bar t} T^{\bar t\bar t}  = -\partial_{\bar a} T^{\bar a \bar t}$, and integrate by parts. The boundary term vanishes and we are left with $\int d^3\bar x \, T^{\bar a \bar t} = 0$. In our numerical simulations,  $J_2 \sim 10^{-27} M^2$. 
\item[$J_3$:] This term is very small because the integral vanishes by the center of mass condition. In our numerical simulations,  
$J_3 \sim 10^{-27} M^2$.
\item[$J_4$:] We identify this term as the spin angular momentum, that is, the angular momentum of the body about its center of mass. To see this, we first compute $\nabla_\mu \xi^\nu_{(\phi)}$ using Schwarzschild coordinates $t$, $r$, $\theta$ and $\phi$.
Let the orbit lie in the equatorial plane and define $\rho_1 = 1/\sqrt{f}$ (so that $f = N^2 = 1 - 2M/r$), $\rho_2 = r$ and $\rho_3 = r \sin\theta$. Since the Killing vector field has components $\xi^\nu_{(\phi)} = \delta^\nu_\phi$, we find that in the 
equatorial plane $\theta = \pi/2$, 
\be
    \nabla_\mu \xi_\nu^{(\phi)} = r\delta_\mu^r \delta_\nu^\phi - r \delta_\mu^\phi  \delta_\nu^r   \ .
\ee
It follows that 
\be
    \nabla_{\bar a} \xi^{(\phi)}_{\bar b} = r\left( e^r_{\bar a} e^\phi_{\bar b} - e^\phi_{\bar a} e^r_{\bar b} \right) \ .
\ee
Let us assume that the triad vectors $e^\mu_{\bar x}$ are 
given by Eqs.~(\ref{orthotetrad}). Then 
\be\label{Daxib}
    \nabla_{\bar x} \xi_{\bar z}^{(\phi)} 
    = -\nabla_{\bar z} \xi_{\bar x}^{(\phi)} 
    = \gamma \sqrt{f}  \ ,
\ee
with all other components vanishing. 
Combining these results, we find 
\be
    J_4 = \gamma  \sqrt{f} \int d^3\bar x \sqrt{\bar g} \left( \bar x \, T^{\bar z\bar t} - \bar z\,  T^{\bar x \bar t} \right)  \ .
\ee
This is, apart from the relativistic factor $\gamma$ and the curved space factors  $\sqrt{f}$ and $\sqrt{\bar g}$,  the spin angular momentum in flat spacetime. (Note that the
triad vector $e^\mu_{\bar y}$ points along the axis of 
rotation; that is, in the $\partial/\partial\theta$ direction.)

A spherical body of mass ${\mathcal M}\sim 10^{-18}M$ and radius ${\mathcal R} \sim 10^{-1}M$ has moment of inertia ${\mathcal I}\approx {\mathcal M}{\mathcal R}^2 \sim 10^{-20} M^3$. In our simulations, after pericenter the elastic body turns through an 
angle of $\Delta\phi \sim 1/2$  in a time of $\Delta t \sim 500M$. Thus, the angular velocity is $\omega \sim 10^{-3}/M$. 
The spin angular momentum is roughly $J_4 \approx {\mathcal I}\omega 
\sim 10^{-23} M^2$. 
\item[$J_5$:] This term is small compared to the spin $J_4$ 
because the acceleration 
term $A_{\bar a} \xi_{\bar b}^{(\phi)}$ is small compared to 
$\nabla_{\bar a}  \xi_{\bar b}^{(\phi)}$. In our numerical 
simulations the acceleration reaches a maximum $A_{\bar a} \sim 10^{-7}/M$ near pericenter.  Given $\xi_{\bar a}^{(\phi)} \sim 10 M$, we find $A_{\bar a} \xi_{\bar b}^{(\phi)} \sim 10^{-6}$. 
Comparing to $J_4$, we find $J_5 \sim 10^{-29} M^2$. 
\item[$J_6$:] The integral in $J_6$ is the second moment of the mass distribution of the elastic body; for example, $\int d^3\bar x \, \bar x^2  T^{\bar t\bar t} \approx 4\pi {\mathcal R}^5 \rho/15 \sim 10^{-21}M^3$. The integral is multiplied by 
curvature terms and an acceleration term. The acceleration 
term yields a relatively small contribution of $\sim 10^{-28}M^2$ to the angular momentum, since   
$A_{\bar a} \sim 10^{-7}/M$ and $\nabla_{\bar a} \xi^{(\phi)}_{\bar b} \sim 1$. The curvature terms are much larger.  The Riemann tensor components are small multiples of $M/r^3$, which is 
$\sim 10^{-3}/M^2$ at pericenter. The components 
$\xi^{(\phi)}_{\bar t}$ and $\xi^{(\phi)}_{\bar a}$ of the Killing 
vector field are 
of order $10 M$ at pericenter. Thus, the curvature terms in 
$J_6$ contribute a factor of  $\sim 10^{-2}/M$ which, combined with the 
second moment of the mass distribution, gives $J_6 \sim 10^{-23} M^2$.

Our simulations show that $J_6$ is somewhat smaller than the estimate of $10^{-23}M^2$. This is because the 
body, even near pericenter, is only modestly distorted from a spherical shape. If we approximate the body as spherical, then the second moment of the mass distribution is proportional to the identity, 
$\int d^3\bar x\, x^{\bar a} x^{\bar b} T^{tt} \propto \delta^{\bar a\bar b}$. The first Riemann term in $J_6$ has the factor $R^{\bar t}{}_{\bar a\bar b\bar t} \delta^{\bar a\bar b} = -R^{\bar t}{}_{\bar t}$, which vanishes 
by the vacuum Einstein equations. Likewise, for the 
second Riemann term, $R^{\bar c}{}_{\bar b\bar c\bar t} \delta^{\bar a\bar b} = -R^{\bar c}{}_{\bar t} = 0$. 
In our simulations $J_6$ reaches a peak value of 
roughly $J_6 \sim 10^{-24}M^2$ as the body passes the black hole. 
\item[$J_7$:] This term is small, $J_7 \sim 10^{-28} M^2$,  because the integral 
is approximately zero. For a uniform body with rigid rotation, 
all odd moments of the momentum distribution will vanish  due to  the symmetry 
$T^{\bar a \bar t}(\bar x,\bar y,\bar z) = -T^{\bar a \bar t}(-\bar x,-\bar y,\bar z)$. As we will see, in our simulations the deformation of the elastic body is primarily in the spherical 
harmonic modes  $(\ell,m) = (0,0)$ and $(2,\pm 2)$. 
These deformations respect this symmetry so the 
second moment of $T^{\bar a \bar t}$ is very small. 
\end{itemize}

\section{Initial data}
\label{sec:ini}

Here we describe the construction of initial data for the numerical evolution of a hyperelastic sphere in a scattering orbit around a Schwarzschild black hole. In flat spacetime and in a frame in which the elastic body is at rest and relaxed, the body is spherical. We find the point which is equidistant from all points on the surface, the geometric center point, $C$. We assume that point $C$ corresponds to the material point with matter space coordinates $\zeta^i=(0,0,0)$. We discretize the matter space assuming that the sphere is relaxed in physical flat spacetime, with the matter space metric equal to the relaxed metric $\epsilon_{ij}$. 

Initially, the sphere is at a large distance from the black hole and moving towards it. We assume that initially,  the spacetime coordinates and velocity of $C$ correspond to those of a parabolic geodesic. For the Schwarzschild geometry, the geodesic equations for motion in the equatorial plane are 
\begin{align}
    \dot r^2 &= {\mathcal E}^2 - 1 + 2M/r - {\mathcal L}^2/r^2 + 2M{\mathcal L}^2/r^3 \ ,\\
    \dot\phi &= {\mathcal L}/r^2 \ , \\
     \dot t &= \frac{{\mathcal E}}{1 - 2M/r} \ ,
\end{align}
where the dot is a derivative with respect to proper time. 
Here, ${\mathcal E}$ and ${\mathcal L}$ are the conserved energy and angular momentum in the orbital plane. 

We choose ${\mathcal E} = 1$ for a parabolic orbit. The angular momentum can be written in terms of pericenter distance $r_p$ as 
\be
    {\mathcal L}^2 = \frac{2Mr_p}{1 - 2M/r_p} \ .
\ee
For a given $r_p$, we evolve the spacetime coordinates and velocity of point $C$ using fourth-order Runge--Kutta for 1000 time steps with a step size of $2.0303\times 10^{-3}\,M$ using the geodesic equations and obtain a Fermi-normal frame carried by point $C$.  
The time-step size is chosen as approximately 1000 
times smaller than the time step used in the numerical evolution of Sec.~\ref{sec:formalism}. With fourth-order 
Runge--Kutta, this reduces 
the truncation error to round-off level. 

A plausible initial condition is that the elastic sphere is in quasistatic equilibrium under elastic and tidal forces in the initial data frame, the Fermi frame attached to the center $C$. 
Love \cite{love1892treatise} has solved the problem of a solid homogeneous elastic sphere in static equilibrium under the action of an external Newtonian potential $\Phi_{\mathrm{ext}}$ that can be expressed in terms of spherical harmonics. The deformation of the sphere is assumed to be small. The bulk static equilibrium equation is
\be
\left({\lambda + \mu} \right)  \nabla^a \nabla_b \xi^b + {\mu} \nabla^b \nabla_b \xi^a - \rho_0 \nabla^a \Phi_\mathrm{ext} = 0 \ , \label{eq:bulkstatic}
\ee 
and the boundary equation is
\be 
\lambda \nabla_a\xi^a n^c + \mu(\nabla^c\xi^d+\nabla^d\xi^c) n_d = 0 \ , \label{eq:bc}
\ee 
where $\xi^a$ is a small displacement from the configuration where the elastic body is relaxed and $\lambda$ and $\mu$ are the Lam\'e elastic parameters.

The Newtonian tidal potential of a black hole of mass $M$ can be expanded in terms of spherical harmonics. Assuming that the geometric center of the sphere is on the $z$ axis and the central mass is at the origin, the tidal potential can be approximated by 
\be 
\Phi_\mathrm{ext}(R,\Theta) = -\sqrt{{4\pi}/{5}} ({M}/{r^3}) R^2 Y_{20}(\Theta) \ , \label{eq:phiext}
\ee
where $r$ is the areal radius of the geometric center of the elastic sphere. Also,  $R$ and $\Theta$ are the radial distance and azimuthal angle of an element of the elastic body from the geometric center of the body. The solution $\vec{\xi}(R,\Theta)$ satisfying Eqs.~(\ref{eq:bulkstatic}) and (\ref{eq:bc}) for $\Phi_{\mathrm{ext}}(R,\Theta)$ in Eq.~(\ref{eq:phiext}) is obtained from Section 181 of Ref.~\cite{love1892treatise} by setting $g=0$ and $\omega=0$ for zero self-gravity and for the sphere to be non-rotating. One can verify that the following solution satisfies the bulk and boundary equations, 
\be 
\vec{\xi}(R,\Theta) = f(R) Y_{20}(\Theta) {\hat R} + g(R) \frac{\pd Y_{20}}{\pd \Theta} {\hat \Theta} \ ,\label{eqn:xistatic}
\ee 
where 
\bse 
\begin{align}
f(R) &= \frac{k \rho_0}{70(\lambda + 2\mu)}\bigg[ -20R^3 + 14a^2R \frac{(\lambda+ 3\mu)}{\mu}  \nono \\
&+\frac{(7\lambda+6\mu)}{\mu(19\lambda+14\mu)}  \bigg( 42R(a^2-R^2)(\lambda+\mu) \nono\\
&\qq\qq\qq\qq\qq+ 6R^3(2\lambda+7\mu) \bigg) \bigg] \ ,
\end{align}
\begin{align}
g(R) &= \frac{k \rho_0}{70(\lambda + 2\mu)}\bigg[ -5R^3 + 7a^2R \frac{(\lambda+3\mu)}{\mu} \nono \\
& +  \frac{(7\lambda+6\mu)}{\mu(19\lambda+14\mu)}  \bigg( 21R(a^2-R^2)(\lambda+\mu) \nono\\
&\qq\qq\qq\qq\qq- 2 R^3(2\lambda+7\mu) \bigg) \bigg] \ ,
\end{align} 
\ese
and $k=\sqrt{{4\pi}/{5}} ({M}/{r^3})$. Here, $a$ is the undeformed radius of the sphere.

We compute $\vec{\xi}(R,\Theta)$ at each value of areal 
radius $r$ 
along the (approximately) parabolic geodesic.  We rotate the displacements so that they correspond to  the geometric center of the sphere being located on the $xy$ plane at angle $\phi$ instead of on the $z$ axis. 

The spatial Fermi-normal coordinates of each node are identified as follows. First, we interpret $R$ and $\Theta$ 
as spherical coordinates in the initial data Fermi frame; that is, $R = \sqrt{(\bar x^1)^2 + (\bar x^2)^2 + (\bar x^3)^2}$ and $\Theta = \arctan(\sqrt{(\bar x^1)^2 + (\bar x^2)^2)} / \bar x^3)$. We then express the solution of Eq.~(\ref{eqn:xistatic}) as a vector $\xi^{\bar a}(\bar x)$ in the Fermi frame. Assuming 
the matter space coordinates $\zeta^i$ are Cartesian 
with relaxed metric $\epsilon_{ij}$, we identify 
the matter space  with the Fermi 
frame by $\zeta^i \leftrightarrow x^{\bar a}$. Then the 
matter space node  $\zeta^i$ is placed in the initial data Fermi frame by  
\be
    \zeta^i \to x^{\bar a} + \xi^{\bar a}(\bar x)  \ .
\ee
Here, the displacement vector $\xi^{\bar a}$ is evaluated 
at the point with coordinate values $x^{\bar a}$ equal to   $\zeta^i$.

\begin{figure}
\begin{center}
\includegraphics{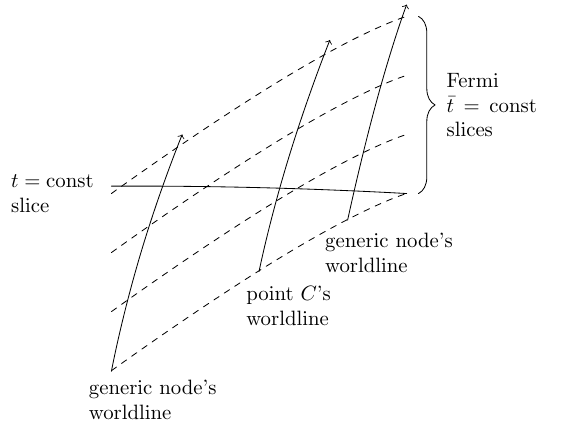}
\caption{\label{fig:initialtimeslices}A $t = \mathrm{const}$ slice that crosses the worldlines of the generic nodes.}
\end{center}
\end{figure}

Having found the Fermi coordinates of each node in the elastic body for the first 1000 (small)  time steps 
of a geodesic orbit,  we convert these to spacetime coordinates using the relations (\ref{FermiSpTmCoordTransf}). 
We then find a $t=\mathrm{const}$ spacetime slice that crosses the worldlines of all nodes, as shown in 
Fig.~\ref{fig:initialtimeslices}. The spacetime coordinates and velocities on this slice are obtained by interpolation and finite differencing using the closest known  coordinate values. These values are used as the initial coordinates and velocities for the spacetime evolution.

\section{Results from tidal encounters}
\label{sec:results}

In this section, we describe the results of the simulation of the hyperelastic sphere for three close encounter scattering orbits around a Schwarzschild black hole. Dimensional quantities are expressed in terms of the black hole mass, $M$.

The equations of motion for a hyperelastic sphere in Schwarzschild spacetime depend on four dimensionless parameters: the longitudinal and transverse sound speeds in the elastic body,
\be
C_L = \sqrt{\frac{\lambda + 2 \mu}{\rho_0}} \ , \qq C_T = \sqrt{\frac{ \mu}{\rho_0}} \ ,    
\ee
the radius of the sphere, $\tilde a = a/M$, and the pericenter distance for a chosen parabolic geodesic that we use to obtain initial data, $\tilde r_p = r_p/M$. We set the values of the sound speeds to $C_L = 0.01$ and $C_T = C_L/\sqrt{3}$.

We create a sphere model of radius $\tilde a = 0.1$ in MATLAB and use MATLAB's mesh generation algorithm \cite{pdetool} to create a linear tetrahedral mesh for a specified maximum edge length of the elements, $h_{\rm max}$, for the sphere model. MATLAB's mesh algorithm enforces that $h_{\rm min}$, the minimum edge length, is equal to $h_{\rm max}/2$. We use four mesh refinements with $h_{\mathrm{max}} = [a/4,a/8,a/16,a/32]$. As the mesh is refined, the total volume of tetrahedral elements converges to $\tilde V_\mathrm{conv}$ and we find the converged radius using the converged volume, $\tilde a_{\mathrm{conv}}= \sqrt[3]{3 \tilde V_{\mathrm{conv}}/4\pi} \approx \tilde a = 0.09976$. We use $\tilde a_\mathrm{conv}$ as the undeformed radius of the sphere. Unless otherwise noted, all plots present results obtained with the finest mesh refinement.

\begin{figure}
\centering
 \includegraphics{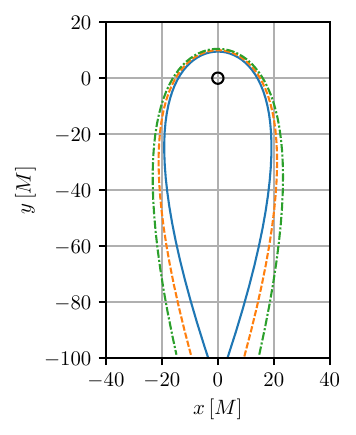}
\caption{Schwarzschild coordinates of the center of mass of the hyperelastic sphere for $\tilde r_p = 9.5, 10,$ and $10.5$. The orbits are counterclockwise in the equatorial plane.
\label{fig:orbits}}
\end{figure}

We investigate the effect of closer encounters by simulating three orbits with pericenter distances $\tilde r_p = 9.5, 10,$ and $10.5$. These orbits are shown in Fig.~\ref{fig:orbits}. At the initial time, the geometric center of the sphere is at an areal radius of $100\,M$. 

For stars, the strength of a tidal encounter is characterized by a dimensionless parameter defined as the ratio of the duration of pericenter passage to the hydrodynamic time of the star \cite{1977ApJ...213..183P,PhysRevD.87.104010}. For an elastic sphere, we define the strength of the encounter as the ratio 
\be         
\eta = \left(\sqrt{{r_p^3}/{M }}\right) / T_{0 2, \rm analytic} \ ,
\ee 
where $\sqrt{r_p^3/M}$ is the 
duration of pericenter passage and $T_{0 2, \rm analytic}$ is the analytically computed period of the spheroidal $n=0, \ell = 2$ normal mode oscillation of the free elastic sphere \cite{JBE23}. Smaller values of $\eta$ correspond to stronger encounters. 

The three orbits with $\tilde r_p = 9.5$, $10$ and $10.5$ correspond to $\eta = 0.71$, $0.77$ and  $0.83$, respectively.  
Thus, for $\tilde r_p = 9.5$, the duration of pericenter passage is about 71\% of an oscillation period. 

\subsection{Deformation of the sphere}

\begin{figure*}   
\centering 
 \includegraphics{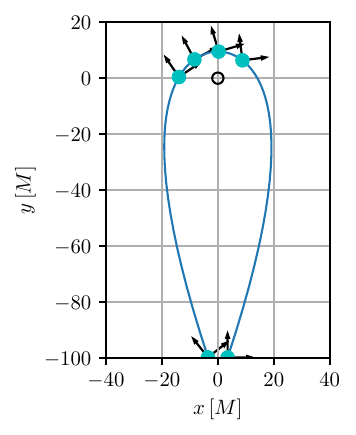}  
 \includegraphics{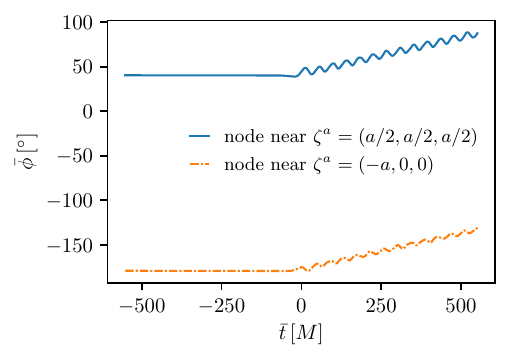}
 \includegraphics{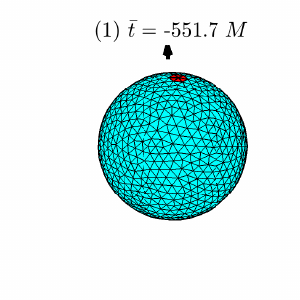}  
 \includegraphics{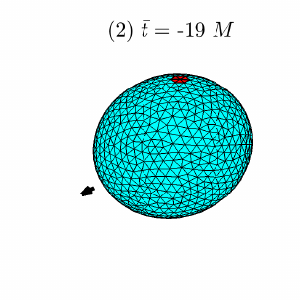}  
 \includegraphics{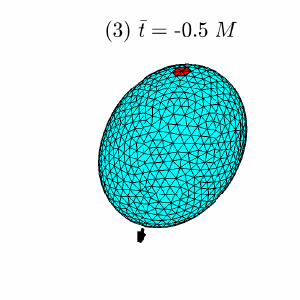}  
 \includegraphics{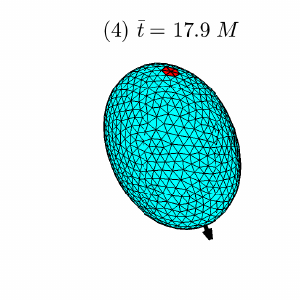}  
 \includegraphics{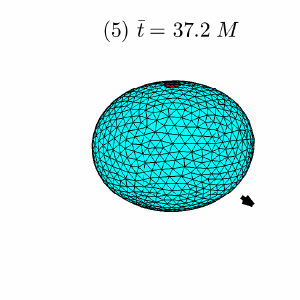}  
 \includegraphics{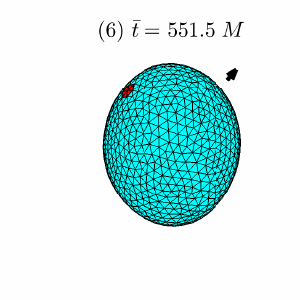}  
\caption{These plots correspond to the $\tilde r_p = 9.5$ orbit and the $h_\mathrm{max}=a/8$ mesh. Top left: points along the orbit at which we plot snapshots of the spatial Fermi coordinates of the sphere. The arrows show the direction of the spatial legs of the fiducial Fermi frame tetrad. They rotate with respect to the black hole frame as they are Fermi-Walker transported. Top right: angle $\bar \phi$ of two nodes in the fiducial Fermi frame which drifts after the encounter. Snapshots $(1)$ to $(6)$: top view in the Fermi frame corresponding to the successive points marked along the orbit. Snapshots $(2)$ to $(5)$ are at constant intervals of $t$. The period of $n=0, \ell=2$ free spheroidal normal modes is $41.2 M$ \cite{JBE23}. Between $(5)$ and $(6)$, the sphere undergoes about 12 oscillations.  The red patch of cells on the surface does not return to its original angle $\bar \phi$ after oscillations. The arrow shows the direction to the black hole. On snapshot $(3)$ the direction of the bulge on the sphere can be seen lagging the direction of the tidal field.
\label{fig:deformation}}
\end{figure*}

We visualize the deformation of the sphere by plotting its spatial Fermi coordinates in the fiducial Fermi frame. An animation of this deformation along the orbit is provided in Supplemental Material~\cite{SupplementalMaterial}. We also analyzed the deformations in the center of mass Fermi frame, but found no visible difference between the two Fermi frames. 
The fiducial Fermi frame is initially aligned with the black hole frame but slowly rotates with respect to the latter as it is Fermi-Walker transported along the orbit. This is seen in the upper-left panel of  Fig.~\ref{fig:deformation}.

Figure~\ref{fig:deformation} shows six snapshots of the sphere at different points along the orbit for the $\tilde r_p = 9.5$ orbit.  
As the sphere approaches the black hole, it is stretched along one direction and compressed in the transverse directions. 
The direction of stretch lags behind the direction of the tidal gravitational field. 
This is seen most clearly in snapshot (3) of Fig.~\ref{fig:deformation}. 

As the elastic body moves away from the black hole, it oscillates and spins. 
The snapshots show a patch of red-colored cells on the surface of the sphere. Observe that 
the colored patch does not 
return to its original angle  in the  Fermi frame, indicating a net rotation. The upper-right 
panel of Fig.~\ref{fig:deformation} shows the angle $\bar\phi$  between the fiducial Fermi $\bar x$--axis  and the direction of two different nodes.  The angles for these nodes (and all other nodes) increase at a similar rate after the encounter.

\subsection{Deviation of the center of mass}

\begin{figure}
\centering
 \includegraphics{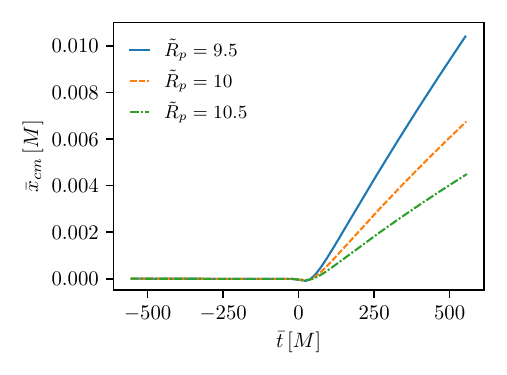}
 \includegraphics{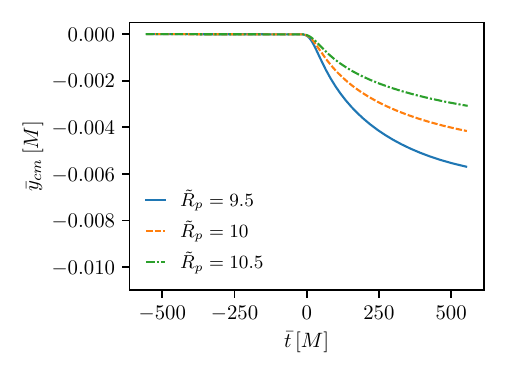}
\caption{Coordinates of the center of mass of the sphere in the geodesic Fermi frame for $\tilde r_p = 9.5, 10,$ and $10.5$. With $\tilde r_p = 9.5$, the center of mass deviation in the $\bar x $ direction is about one tenth of the sphere's radius at the end of the simulation.
\label{fig:comx}}
\end{figure}

As the extended body approaches the black hole, it is deformed, and its second mass moment couples to the octupole tide causing the center of mass to deviate from a geodesic \cite{PhysRevD.87.104010}. We observe that the center of mass, computed using the procedure described in Sec.~\ref{sec:localestcom}, deviates in the $\bar x$--$\bar y$ plane of the geodesic Fermi frame (see Fig.~\ref{fig:comx}).  The center of mass deviation at the end of the simulation in the $\bar{x}$ direction, for $\tilde r_p = 9.5$, $10$, and $10.5$, is approximately $0.0104\,M$, $0.0067\,M$, and $0.0045\,M$, respectively, while the deviation in the $\bar{y}$ direction is approximately $-0.0057\,M$, $-0.0042\,M$, and $-0.0031\,M$. An empirical fit of these data points to a power of $1/r_p$ suggests that the final deviations scale steeply with pericenter distance, approximately as $(1/\tilde{r}_p)^{8}$ in the $\bar{x}$ direction and $(1/\tilde{r}_p)^{6}$ in the $\bar{y}$ direction.

\subsection{Energy}

\begin{figure*}
\includegraphics{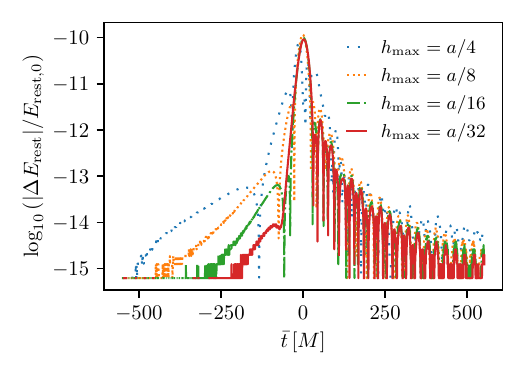}
\includegraphics{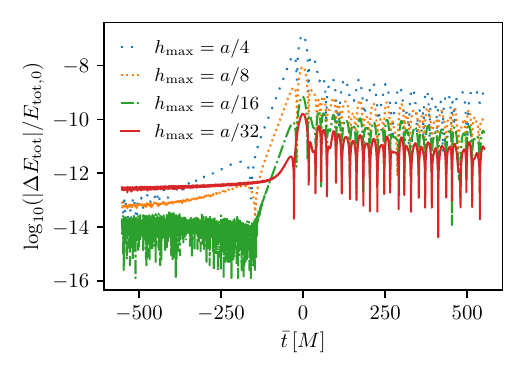}
\caption{Numerical conservation of the rest energy and total energy for the $\tilde r_p = 9.5$ orbit across four mesh refinements  $h_\mathrm{max}$. Left: The fractional change in the rest energy, defined as the deviation from its initial value divided by the initial value. Right: The fractional change in $ E_\mathrm{tot}$. As the mesh is refined, the numerical conservation of the total energy improves approximately as $(1/h_\mathrm{max})^2$. The total energy is conserved to within 9.8 orders of magnitude for the mesh refinement $h_\mathrm{max} = a/32$. Errors in $ E_{\mathrm{tot}}$ that are less than $10^{-12}$ are dominated by floating-point round-off.
\label{fig:ErestandEtot}}
\end{figure*}

We first compute the theoretically conserved rest energy (\ref{eq:Erest}) in the center of mass Fermi frame as described in Sec.~\ref{sec:locestphy}. We find that the rest energy is numerically conserved to 10 orders of magnitude close to pericenter ($-50\,M \lesssim \bar t \lesssim 50\,M$) and to 12 or more orders of magnitude further from the black hole (see Fig.~\ref{fig:ErestandEtot}). We suspect that the larger errors near 
pericenter are due to the neglect of 
higher-order terms in $\bar x^a$ in computing the Fermi quantities. 

Next, we compute the theoretically conserved total energy (\ref{eq:Etot}), which displays numerical conservation to within 9.8 orders of magnitude for the mesh refinement $h_\mathrm{max} = a/32$ (see Fig.~\ref{fig:ErestandEtot}).

\begin{figure}
 \includegraphics{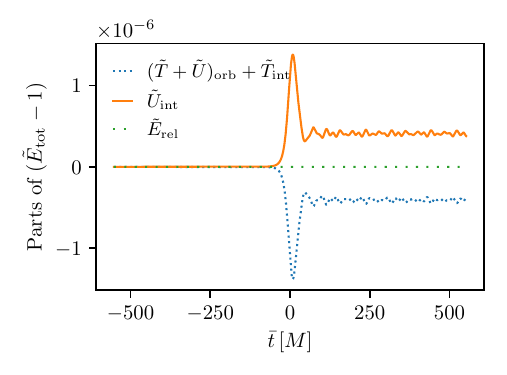}
\caption{The three parts of $(E_{\mathrm{tot}}-E_{\mathrm{rest}})$ (\ref{eq:E3parts}) for the $\tilde r_p = 9.5$ orbit. The tilde denotes specific energies obtained by dividing by the initial rest energy for the $h_\mathrm{max}=a/32$ mesh. The changes in $(\tilde T + \tilde U)_{\rm orb} + \tilde T_{\rm int}$ and $\tilde U_{\rm int}$ lie well above the numerical error floor of $10^{-9.8}$ in $\tilde E_{\rm tot}$ (see Fig.~\ref{fig:ErestandEtot}), while $\tilde E_{\rm rel}$ remains below that floor. The sphere has a constant deformation with a bulge rotating in the $xy$ plane after the encounter explaining the average positive offset of the internal potential energy, $U_\mathrm{int}$.
\label{fig:E3parts}}
\end{figure}

We subtract the rest energy from the total energy and separate it into three parts (\ref{eq:E3parts}); the sum of the kinetic and potential orbital energy and the internal kinetic energy, $(T + U)_\mathrm{orb} + T_\mathrm{int}$, the internal potential energy, $U_\mathrm{int}$, and the contribution of the spatial stress to the energy $E_\mathrm{rel}$. Figure~\ref{fig:E3parts} shows these three parts computed in the center of mass Fermi frame. The decrease in $(T + U)_\mathrm{orb} + T_\mathrm{int}$ is accompanied by an increase in the internal potential energy, $U_\mathrm{int}$, of the elastic sphere. The internal potential energy, $U_\mathrm{int}$, is initially close to zero. As the elastic sphere is tidally deformed, it increases to a maximum value, then decreases, and settles into an average positive offset from zero as the sphere moves away from the black hole. As the elastic body moves away from the black hole, it has an ellipsoidal shape, with the bulge rotating counterclockwise in the $xy$ plane. The constant deformation explains the average positive offset from zero of $U_\mathrm{int}$. The contribution of spatial stress to the energy, $E_\mathrm{rel}$, is found to be very small. 

\begin{figure*}
 \includegraphics{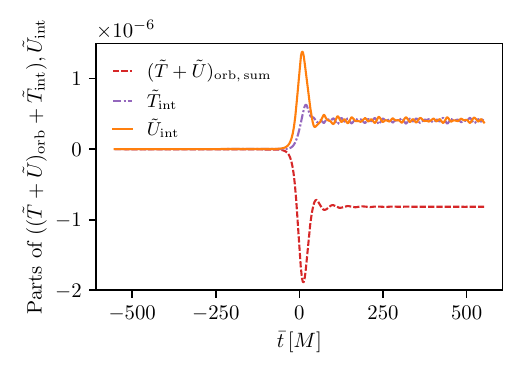}
 \includegraphics{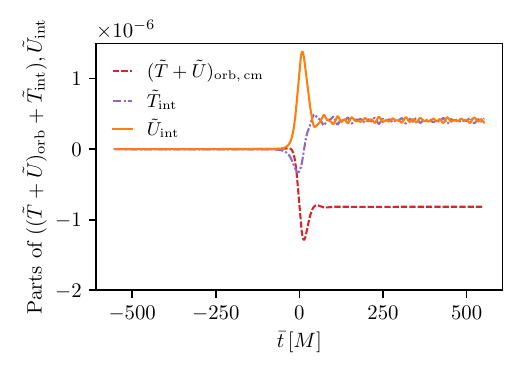}
 \includegraphics{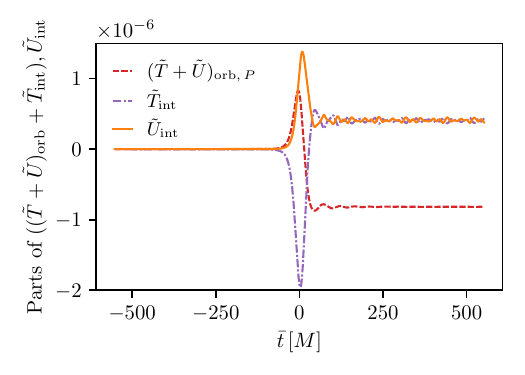}
 \includegraphics{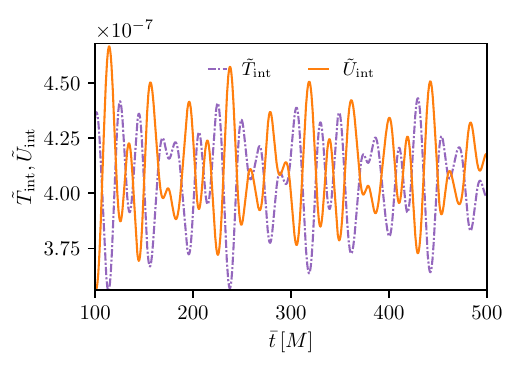}
\caption{Decomposition of $(T + U)_\mathrm{orb} + T_\mathrm{int}$ using three different definitions of $(T + U)_\mathrm{orb}$, and $U_\mathrm{int}$, for the $\tilde r_p = 9.5$ orbit. The tilde denotes specific energies obtained by dividing by the initial rest energy for the $h_\mathrm{max}=a/32$ mesh. Top left: first way of computing the orbital energy as $(T + U)_{\mathrm{orb}, \,\mathrm{sum}}$. Top right: second way of computing the orbital energy as $(T+ U)_{\mathrm{orb}, \, \mathrm{cm}}$. Bottom left: third way of computing the orbital energy as $(T+ U)_{\mathrm{orb}, \, P}$. Bottom right: zoom in on the small‐amplitude oscillations in the top left plot for $\bar t > 100\,M$. The three different ways of computing the orbital energy lead to very different values of the internal kinetic energy, $\tilde T_\mathrm{int}$, close to pericenter ($-70\,M \lesssim \bar t \lesssim 70\,M$). Only the first method, using $(T + U)_{\mathrm{orb}, \,\mathrm{sum}}$, leads to a value of $\tilde T_\mathrm{int}$ that is always positive. In the zoomed plot, we observe that the internal kinetic and potential energy have equal and opposite oscillations about the same average value for $\bar t \gtrsim 200\,M$.
\label{fig:E3partsandEorb3parts}}
\end{figure*}

We further separate $(T + U)_\mathrm{orb} + T_\mathrm{int}$ as described in Sec.~\ref{sec:locestphy} into orbital and internal kinetic energy. Figure~\ref{fig:E3partsandEorb3parts} shows the results of the three ways we considered of carrying out the separation by identifying orbital energy, as $(T + U)_{\mathrm{orb}, \,\mathrm{sum}}$ using Eq.~(\ref{eq:Eorb1}), $(T+ U)_{\mathrm{orb}, \, \mathrm{cm}} $ using Eq.~(\ref{eq:Eorb2}) and $(T+ U)_{\mathrm{orb}, \, P}$ using Eq.~(\ref{eq:Eorb3}). All three ways agree farther away from pericenter. Farther away they show that the resulting internal kinetic energy of the sphere has an average positive shift equal to that of the internal potential energy. The nearly constant kinetic energy that fails to drop to zero becomes clear when we decompose the sphere’s deformation into normal modes: the oscillation is dominated by the $n=0,\ell=2,m=-2$ and $n=0,\ell=2,m=2$ modes, which have almost equal amplitudes and are offset by $90^\circ$. 

For the Saint Venant-Kirchhoff model the internal potential energy density is quadratic in the Lagrange strain \cite{JBE23}. The virial theorem then predicts that the time-averaged internal kinetic energy equals the time-averaged internal potential energy. We observe that the internal potential energy of the sphere has about the same positive shift as the internal kinetic energy. The small and opposite oscillations in the internal kinetic and potential energies visible on the plots are perhaps due to the amplitude of the dominant modes being slightly different.

Using the average change in angle $\bar \phi$ of the nodes with $\bar t$ in the fiducial Fermi frame after the encounter, we compute the mean angular frequency and estimate the specific Newtonian rotational kinetic energy for the $\tilde r_p=9.5$ orbit to be around $4\times 10^{-9}$. Therefore, the rotational kinetic energy contributes negligibly to the observed internal kinetic energy.

The increase in the internal kinetic and potential energy of the sphere is at the expense of its orbital energy, which becomes negative. The orbit changes from an unbound parabolic orbit to a closed orbit as a result of the encounter.

The three ways of computing the orbital energy lead to different computed values of the internal kinetic energy close to pericenter. The method that leads to an internal energy of the elastic sphere that is always positive is using $(T + U)_{\mathrm{orb}, \,\mathrm{sum}}$. Using $(T+ U)_{\mathrm{orb}, \, \mathrm{cm}}$, the orbital energy displays fewer oscillations around $\bar t \approx 70\,M$. 

\subsection{Angular momentum}

\begin{figure}
 \includegraphics{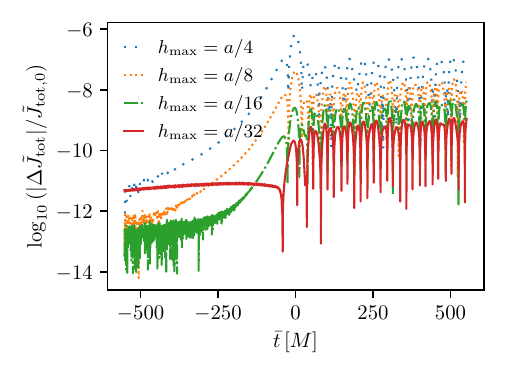}
\caption{Numerical conservation of the total angular momentum for the $\tilde r_p = 9.5$ orbit across four mesh refinements $h_\mathrm{max}$. The tilde in $\tilde J_\mathrm{tot}$ denotes the specific total angular momentum obtained by dividing by the initial rest energy for the $h_\mathrm{max}=a/32$ mesh. We plot the fractional change in $\tilde J_\mathrm{tot}$. As the mesh is refined, the numerical conservation of the total angular momentum improves approximately as $(1/h_\mathrm{max})^2$. The total angular momentum is conserved to 9 orders of magnitude for the mesh refinement $h_\mathrm{max} = a/32$. The error in $\tilde J_{\mathrm{tot}}$ less than $10^{-11}$ is dominated by floating-point round-off.
\label{fig:Jz}}
\end{figure}

We compute the theoretically conserved total angular momentum, $J_{\rm tot}$, (\ref{eq:Jz}) in the center of mass Fermi frame as described in Sec.~\ref{sec:locestphy}. Figure~\ref{fig:Jz} shows that the total angular momentum is numerically conserved to 9 orders of magnitude for the highest-resolution mesh.

\begin{figure}
 \includegraphics{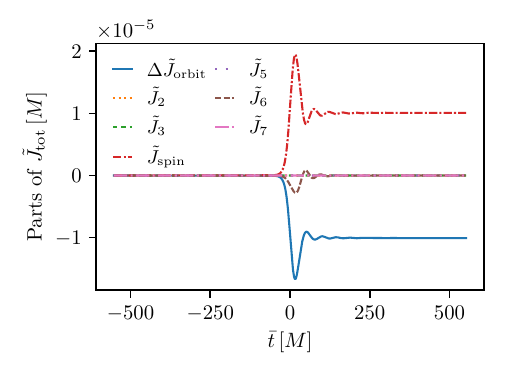}
\caption{
The seven parts of $J_{\mathrm{tot}}$ (\ref{eq:Qrotparts}) for the $\tilde r_p = 9.5$ orbit. The tilde denotes specific angular momenta obtained by dividing by the initial rest energy for the $h_\mathrm{max}=a/32$ mesh. For $\tilde J_\mathrm{orbit}$, we plot the change from the initial value. Changes in $\Delta \tilde J_{\rm orbit}$, $\tilde J_{\rm spin}$ and $\tilde J_6$ lie well above the numerical error floor of $10^{-9}$ in $\tilde J_{\rm tot}$ (see Fig.~\ref{fig:Jz}).  All other terms are smaller than $2.6 \times 10^{-9}\,M$, comparable to the round-off error of the simulation, and are included in the plot but visually overlap with the flat line. Due to the encounter, part of the orbital angular momentum of the sphere is converted to spin angular momentum.
\label{fig:Jzparts}}
\end{figure}

Figure~\ref{fig:Jzparts} shows the seven parts of $J_{\rm tot}$ (\ref{eq:Qrotparts}) for the $\tilde r_p=9.5$ orbit, computed in the center of mass Fermi frame as described in Sec.~\ref{sec:locestphy}. Our estimates of each part (see Sec.~\ref{sec:locestphy}) match the observed values. The spin of the elastic sphere, $J_\mathrm{spin}$, is initially  zero to within numerical round-off error, increases rapidly as it moves to the pericenter, then decreases rapidly and settles on a constant value as the sphere moves away from the black hole. The change in spin is due to the misalignment of the direction of the deformation of the sphere and the direction to the black hole \cite{PhysRevD.87.104010} (see snapshot (3) in Fig.~\ref{fig:deformation}), which results in a torque on the sphere. As a result of the encounter with the Schwarzschild black hole, the elastic body loses orbital angular momentum while gaining spin angular momentum.

\subsection{Decomposition into normal modes}

In snapshot (5) of Figure~\ref{fig:deformation}, the ellipsoid’s longest principal axis measures approximately \(0.215\,M\).  Relative to the sphere’s undeformed radius, this corresponds to a strain of about \(7.8\%\).  At this level of deformation, we therefore anticipate observable nonlinear contributions when we analyze the sphere’s oscillations using the small-deformation normal-mode framework.

In the fiducial Fermi frame, the sphere’s average rotation frequency is approximately $0.009$ times the nonrelativistic $n=0,\ell=2$ analytical spheroidal mode oscillation frequency \cite{JBE23}. Given this slow rotation rate, we expect Coriolis and centrifugal forces to be negligible. However, when tracking a node’s displacement from its initial equilibrium position in the fiducial Fermi frame, we observe the displacement growing over time due to rotation. To correct for this drift, we adopt a Fermi frame that co-rotates with the elastic sphere. At each time step, we compute the average angular shift $\bar\phi$ in the fiducial frame and rotate the frame by $\bar\phi$ to obtain the rotating Fermi frame.

The displacement field corresponding to the nonrelativistic normal modes of a free solid elastic sphere, which can be classified into spheroidal and torsional modes and is described in Ref.~\cite{JBE23}, can be used as a basis to express a generic spatial displacement field. We focus on the spheroidal modes. The observed displacements of the nodes in the rotating fiducial frame at time $\bar t$ can be expanded as
\be\label{expansionofdisplacement}
\vec{\xi} = \sum_{n  \ell m} C_{n  \ell m} \; \vec{ \Xi}_{n  \ell m}(R, \Theta, \Phi) + ... \ ,
\ee
where the displacement field for spheroidal modes, $\vec{ \Xi}_{n  \ell m}$, are given in Eq.~(62) of Ref.~\cite{JBE23}. (The unwritten terms 
in Eq.~(\ref{expansionofdisplacement}) represent the torsional modes.)  We normalize $\vec{ \Xi}_{n  \ell m}$ such that
\be 
\int_S d^3\zeta \sqrt{\epsilon} \; \vec{\Xi}_{n  \ell m} \cdot \vec{\Xi}_{n'l'm'}  =  \delta_{nn'}\delta_{ll'} \delta_{mm'} \ .
\ee
We numerically evaluate the following integral to obtain the decomposition coefficient $C_{n  \ell m}$, for $n=0,1,2$, $\ell=0,1,2,3$ and $m=-\ell, \ldots, \ell$ at each value of Fermi time $\bar t$,
\be
C_{n  \ell m}  = \int_S d^3\zeta \sqrt{\epsilon} \; \vec{\xi} \cdot \vec{\Xi}_{n  \ell m} \ .
\ee 

Away from pericenter, the coefficients $C_{n\ell m}$ for most modes vary sinusoidally in $\bar t$, at frequencies very close to the analytical spheroidal normal mode frequency $f_{n \ell, \rm analytic}$ of the free elastic sphere. Numerical convergence is confirmed by evaluating $C_{n\ell m}$ across four levels of mesh refinement.

\subsubsection{Fundamental modes}

\begin{figure*}
\includegraphics{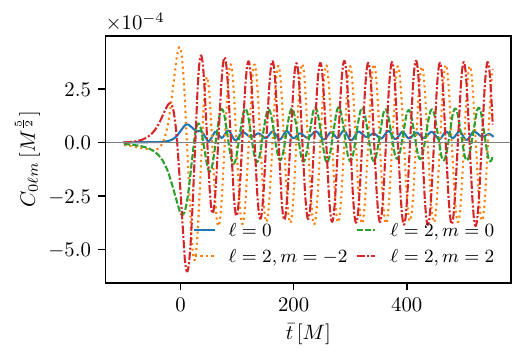}
\includegraphics{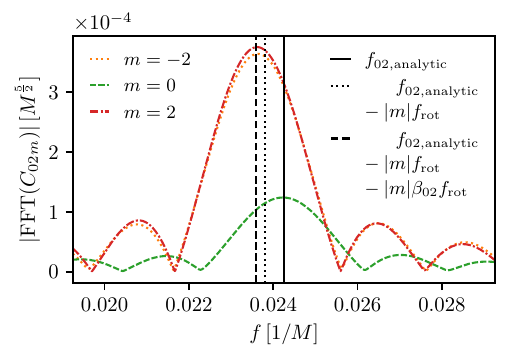}
\caption{Fundamental modes excited in the elastic sphere for the $\tilde r_p = 9.5$ orbit. Left: time-series of the decomposition coefficients for the four highest-energy modes in the rotating fiducial Fermi frame. Right: Fast Fourier Transform (FFT) of the $n=0, \ell=2, m=\pm 2, 0$ modes for $\bar t > 50\,M$. The solid vertical line marks the $n=0, \ell=2$ analytical frequency of the spheroidal normal mode of the free elastic sphere. The dashed  vertical line marks the analytical frequency of $\ell=2, m=\pm 2$ modes which are shifted due to the rotation of the frame and the first-order effect of the rotation of the elastic body.
\label{fig:n0modes}}
\end{figure*}

The dominant modes excited in the elastic sphere are the $n=0, \ell=2, m=\pm 2$ modes (see Fig.~\ref{fig:n0modes}), which have a $90^\circ$ phase difference corresponding to circular polarization which rotates counterclockwise. In addition to the $m=\pm2$ pair, the $\ell=2,m=0$ mode is also excited at less than half the amplitude of the $m=\pm2$ modes. 

Although the tidal field contains no $\ell=0$ component, the large ellipsoidal deformation near pericenter induces a net radial compression in the elastic sphere, thereby exciting a small-amplitude breathing ($n=0,\ell=0$) mode. The $n=0,\ell=0$ and $n=0,\ell=2,m=0$ modes oscillate about a positive offset. The value $C_{000}>0$ signifies a radial compression, and $C_{020}>0$ a prolate distortion.  These offsets may arise because the $n=0,\ell=2,m=\pm2$ quadrupole deformation has a nonzero time average and this large deformation leads to a steady radial compression and axial elongation.

In Fig.~\ref{fig:n0modes} we show the results of 
a Fast Fourier Transform (FFT) applied to the time-series for  the decomposition coefficients $C_{02-2}$, $C_{020}$ and $C_{022}$. For $C_{020}$, the FFT peak lines up with the $n=0, \ell=2$ spheroidal normal mode frequency of the free elastic sphere, $f_{02, \rm analytic}$. For 
$C_{02\pm 2}$, we must account for the effects of rotation on the 
normal mode frequency.  
For the elastic sphere this has been analyzed using perturbation theory in Ref.~\cite{Backus1961}. The first-order perturbation shift in frequency is $-|m|\beta_{n\ell} f_{\rm rot}$, where $ f_{\rm rot}$ is the 
rotation frequency of the sphere and $\beta_{n\ell}$ is given in Eq.~(35) in that reference. For our sphere parameters, we find $\beta_{02}=0.452396$.
We must also include a correction $-|m|f_{\rm rot}$ to the frequency due 
to the rotation of the Fermi frame. Together, these corrections predict 
a frequency for the $n=0, \ell=2, m\neq 0$ modes 
of $f_{02, \rm analytic} - 2f_{\rm rot} - 2\beta_{02}f_{\rm rot}$
As shown in Fig.~\ref{fig:n0modes}, the FFT peaks for $C_{02-2}$ and $C_{022}$ agree with $f_{02, \rm analytic} - 2f_{\rm rot} - 2\beta_{02}f_{\rm rot}$. 

\subsubsection{First and second overtone modes}

\begin{figure*}
\includegraphics{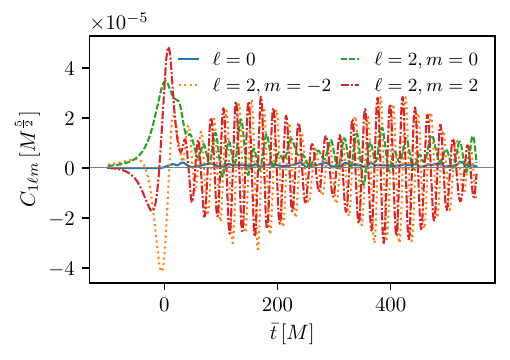}
\includegraphics{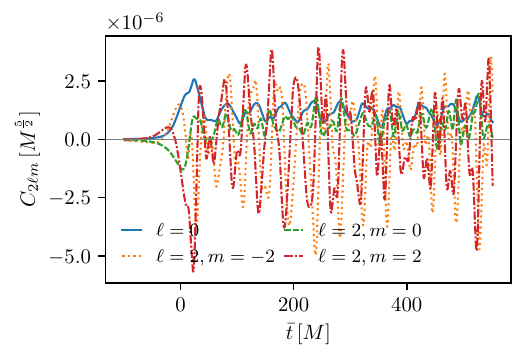}
\caption{Time-series of the decomposition coefficients for overtones for the $\tilde r_p = 9.5$ orbit. Left: first overtones. The $n = 1, \ell = 2, m = \pm 2$ oscillations exhibit a distinct envelope pattern which might be caused by nonlinear coupling between the $n=0, \ell=2, m=0$ and $n=1, \ell=2, m=\pm 2$ modes. Right: second overtones.   The time-series for the decomposition coefficients of the second overtones have frequencies approximately equal to the $n=0, \ell=2$, analytical spheroidal normal mode frequency of the free elastic sphere, which indicates that they result from the nonlinear deformation of the sphere.
\label{fig:n1modes}}
\end{figure*}

The first overtones, $n=1, \ell=2, m=0,\pm 2$, are excited with amplitudes about an order of magnitude smaller than the fundamental modes, $n=0, \ell=2, m=\pm 2$.  The overtones are shown in Fig.~\ref{fig:n1modes}. From the time-series plot, these overtones have frequencies approximately equal to the $n=1, \ell=2$, analytical spheroidal normal mode frequency of the free elastic sphere. 

The decomposition coefficients $C_{02-2}$ and $C_{022}$ exhibit a distinct envelope pattern with a period of about $250M$. We performed a test in flat spacetime, with the nodes of the sphere having initial displacements and velocities corresponding to the normal modes in Fig.~\ref{fig:n1modes}. 
When the initial displacements and velocities include  the $n=0, \ell=2, m=0$ mode,  we observe the same envelope pattern for the $n=1, \ell=2, m=\pm 2$ modes. Without the $n=0, \ell=2, m=0$ mode, the $n=1, \ell=2, m=\pm 2$ modes do not exhibit the envelope. It appears that the 
envelope for the $n=1, \ell=2, m=\pm 2$ modes is caused by nonlinear 
coupling with the $n=0, \ell=2, m=0$  mode. 

The decomposition coefficients for the second overtone modes, $n=2, \ell=2, m=\pm 2$, are roughly two orders of magnitude smaller than the fundamental, $n=0, \ell=2, m=\pm 2$ (see Fig.~\ref{fig:n1modes}). The time-series for the  second overtones have frequencies approximately equal to the $n=0, \ell=2$, analytical spheroidal normal mode frequency of the free elastic sphere. 
This indicates that the second overtones are not directly excited from the tidal encounter. Rather,  the second overtones result from  nonlinear coupling 
to the $n=0$, $\ell=2$ mode. 

\subsubsection{Variation of decomposition coefficients with pericenter distance}

\begin{figure}
 \includegraphics{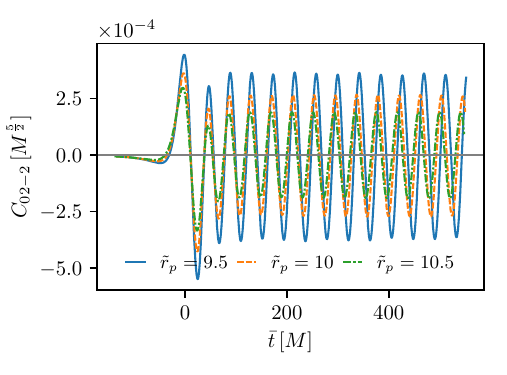}
 \includegraphics{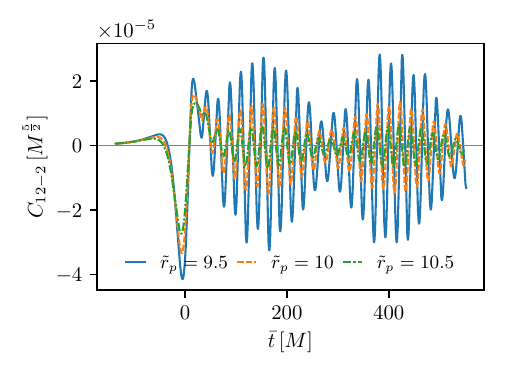}
 \includegraphics{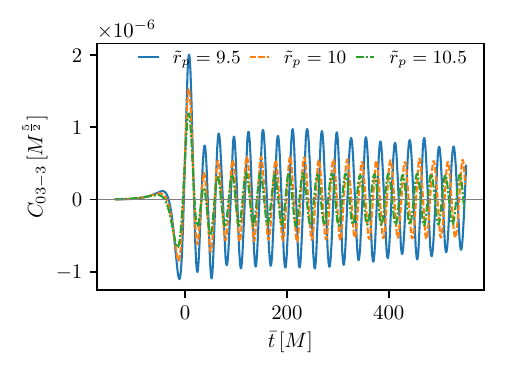}
\caption{Time-series of the decomposition coefficients, $C_{0,2-2}$, $C_{1,2-2}$ and $C_{0,3-3}$, for three pericenter distances, $\tilde r_p = 9.5, 10,$ and $10.5$.
\label{fig:modes3Rp}}
\end{figure}

Figure~\ref{fig:modes3Rp} shows the decomposition coefficients $C_{02-2}$, $C_{12-2}$ and $C_{03-3}$ for $\tilde r_p = 9.5$, $10$, and $10.5$. Each  coefficient varies sinusoidally with approximately the same frequency as the corresponding spheroidal normal-mode frequency of the free elastic sphere $f_{n\ell, \rm analytic}$. The amplitudes of $C_{12-2}$ and $C_{03-3}$
are approximately one and two (respectively) orders of magnitude
smaller than that of $C_{02-2}$. 
As the pericenter distance, $\tilde r_p$, decreases, the amplitudes for $C_{02-2}$ and $C_{03-3}$ increase approximately linearly while the amplitude of $C_{12-2}$ increases more rapidly than linearly.

\section{Conclusions}
\label{sec:concl}

In Ref.~\cite{JBE23}, we proposed a numerical scheme to model the general relativistic motion of extended hyperelastic bodies of any shape in general background spacetimes and validated the code in flat spacetime by comparing the numerical solution to the known analytical normal modes of a free elastic sphere. In this paper, we used our code with the Schwarzschild metric to model the motion of a $0.1\,M$-radius hyperelastic sphere in close encounter orbits around the black hole. The sphere is initially in quasistatic equilibrium with the local $\ell=2$ tidal field to prevent initial oscillations. 

We visualized the deformation of the sphere along its orbit in a Fermi frame carried by the sphere. We used several Fermi frames to compute a local estimate of the center of mass. We computed the theoretically conserved rest energy, total energy and total angular momentum in the Fermi frame along the center of mass worldline and validated the simulation showing their excellent numerical conservation and correct rate of convergence with respect to four mesh refinements. 

We analyzed the total energy and total angular momentum by identifying their distinct components and computed these quantities using the numerical solution. We separated the total energy using the natural separation of the SEM tensor (\ref{semtensorforelastic}) into internal energy, relativistic spatial stress and the remaining part which is a combination of orbital energy and internal kinetic energy. We examined various ways of splitting this last part 
into orbital energy and internal kinetic energy. 
For the total angular momentum, we identify different parts using moments of the SEM tensor. The sphere loses orbital angular momentum in exchange for spin angular momentum as a result of the encounter.  By decomposing the deformation of the sphere into normal modes, we obtained detailed information on the modes excited due to the encounter.

In the future, we plan to use our numerical framework to investigate spin-curvature effects by constructing initial data for the sphere with spin. We are interested in comparing the deviation of the center of mass and changes in spin for scattering orbits with MPD evolutions using different spin supplementary conditions. Furthermore, we are interested in modeling the motion of a solid sphere in Kerr spacetime to study the effects of the black hole's spin on the body. 

\begin{acknowledgments}
N.J. acknowledges support from the NASA Astrophysics Theory Program under award No.~80NSSC22K1898.  C.R.E.~acknowledges 
support from NSF Grant Nos.~PHY-2110335 and PHY-2409604 to 
the University of North Carolina at Chapel Hill. We acknowledge the computing resources provided by North Carolina State University High Performance Computing Services Core Facility (RRID\:SCR\_022168).
\end{acknowledgments}

\appendix
\section{Rest energy current}\label{appendix:conservation}
The rest energy current ${\mathcal J}^\mu = (\sqrt{\epsilon} /\sqrt{f} ) \rho_0 U^\mu$ satisfies $\nabla_\mu {\mathcal J}^\mu = 0$. To prove this, we begin with the observation that  $\lambda$ and $\zeta^i$  can be used collectively as a system of spacetime coordinates. Let $\xi^\mu = (\lambda,\zeta^i)$ denote these coordinates. The 
relation $x^\mu = X^\mu(\lambda,\zeta)$ can be written as $x^\mu = X^\mu(\xi)$, which describes the transformation between the two systems of coordinates. We write the inverse 
transformation as $\xi^\mu = \Xi^\mu(x)$. 

Note that the current, written as ${\mathcal J}^\mu= (\sqrt{\epsilon} /\sqrt{f} ) \rho_0 U^\mu$ is a function of $\xi^\mu$. We can view ${\mathcal J}^\mu$ as a function of $x^\mu$ by evaluating at $\xi^\mu = \Xi^\mu(x)$. 

The tensor transformation rule for the metric gives
\be
    \det\left( \partial x/\partial \xi \right) = \sqrt{-\gamma}/\sqrt{-g}  \ ,
\ee
where $g$ and $\gamma$ are the determinants of the metric tensor components written in 
coordinates $x^\mu$ and $\xi^\mu$,  respectively. 
From  
$x^\mu = X^\mu(\lambda,\zeta)$ we have $dx^\mu = \dot X^\mu d\lambda + X^\mu_{,i} d\zeta^i$. The metric in coordinates $\xi^\mu$ is obtained by substituting the result for $dx^\mu$ into the line element $ds^2 = g_{\mu\nu} dx^\mu dx^\nu$. This yields
\be
    ds^2 = -\alpha^2  d\lambda^2 + 2\alpha v_i d\lambda d\zeta^i 
    + (f_{ij} - v_i v_j) d\zeta^i d\zeta^j \ ,
\ee
where $v_i = U_\mu X^\mu_{,i}$. It follows that the determinant of the spacetime metric in these coordinates is $\gamma = -\alpha^2 f$, where $f$ is the determinant of $f_{ij}$. Thus, we find 
\be
    \det\left( \partial x/\partial \xi \right) = \alpha \sqrt{f}/\sqrt{-g} \ .
\ee
Use this identity to eliminate $\sqrt{f}$ from the rest energy current: 
\be
    {\mathcal J}^\mu = \frac{\sqrt{\epsilon} \rho_0 }{\sqrt{-g} \det(\partial x/\partial \xi) } \dot X^\mu \ .
\ee

Now compute the covariant divergence of $ {\mathcal J}^\mu$. The factor $\sqrt{-g}$ can be brought outside the covariant 
derivative, 
\be
    \nabla_\mu {\mathcal J}^\mu = \frac{1}{\sqrt{-g}} 
    \nabla_\mu \left( \frac{\sqrt{\epsilon} \rho_0}{\det(\partial x/\partial \xi)} \dot X^\mu \right) \ .
\ee
The quantity inside the parentheses is a vector density, so 
we can replace the covariant derivative $\nabla_\mu$ with a coordinate derivative $\partial_\mu$. Applying the product rule, we have 
\begin{align} \label{delJequation}
\sqrt{-g} \nabla_\mu {\mathcal J}^\mu &= 
\dot X^\mu \partial_\mu \left( \sqrt{\epsilon} \rho_0/\det(\partial x/\partial \xi) \right)  \nono\\
&\qquad +  \left( \sqrt{\epsilon} \rho_0/\det(\partial x/\partial \xi) \right) \partial_\mu \dot X^\mu  \ .
\end{align}
Observe that $\dot X^\mu \partial_\mu = \partial/\partial\lambda$ and that $\epsilon$ and $\rho_0$ are independent of $\lambda$. Then the first term above 
is proportional to  the derivative of $\det(\partial x/\partial\xi)$ with respect to $\lambda$. 
We can use the identity $\delta (\det M) = (\det M) {\rm Tr} (M^{-1} \delta M)$, where $M$ is any  matrix and ${\rm Tr}$ denotes the trace. This gives 
\be
    \frac{\partial(\det(\partial x/\partial \xi))}{\partial\lambda} = \det(\partial x/\partial \xi)
    \frac{\partial \Xi^\mu}{\partial x^\nu} \frac{\partial}{\partial \lambda} \left( \frac{\partial X^\nu}{\partial \xi^\mu} \right) \ ,
\ee
which simplifies to 
\be
    \frac{\partial(\det(\partial x/\partial \xi))}{\partial\lambda} = \det(\partial x/\partial \xi) 
    \frac{\partial \dot X^\nu}{\partial x^\nu}
\ee
Using this result in Eq.~(\ref{delJequation}), we see that the first and second terms cancel. Thus, we find $\nabla_\mu {\mathcal J}^\mu = 0$. 

\section{Fermi coordinates}\label{FermiCoordAppendix}
The spacetime coordinates are $x^\mu$, where the surfaces of constant $t \equiv x^0$ are spacelike. Let $x^\mu = X^\mu_{(cw)}(t)$ be a timelike worldline, the ``central worldline," parametrized by $t$. Thus, we have $t = X^0_{(cw)}(t)$. The central worldline need not be a geodesic. 

Let 
\be
    \alpha^{(cw)} = \frac{d\bar t}{dt} =  \sqrt{-\dot X^\mu_{(cw)} \dot X^{(cw)}_\mu}  \label{alphadefappendix}
\ee
where $\bar t$ is proper time along the central worldline and  the dot denotes $d/dt$. As described 
in subsection \ref{ssFermiCoords}, we integrate this 
equation numerically to obtain $\bar t(t)$, the proper time as a function 
of coordinate time $t$.  

The four-velocity and acceleration of the central worldline are
\begin{widetext}
\begin{align}
    U^\mu_{(cw)} &= \dot X^\mu_{(cw)}/\alpha_{(cw)}  \ ,\\
    A^\mu_{(cw)} &= U^\alpha_{(cw)} \nabla_\alpha U^\mu_{(cw)} 
    = \frac{1}{\alpha_{(cw)}^2} (\delta^\mu_\nu + U^\mu_{(cw)} U_\nu^{(cw)})
    \left(\ddot X^\nu_{(cw)} + \Gamma^\nu_{\alpha\beta}  \dot X^\alpha_{(cw)} \dot X^\beta_{(cw)} \right) \ .
\end{align}
\end{widetext}
The Christoffel symbols that appear in the equation for $A^\mu$ 
are functions of $t$, given explicitly by evaluation at $x^\mu = X^\mu_{(cw)}(t)$. That is,  $\Gamma^\nu_{\alpha\beta} \equiv \Gamma^\nu_{\alpha\beta}(X_{(cw)}(t))$. We will 
sometimes denote this as $\Gamma^\mu_{\alpha\beta}(t)$. 

Let $x^\mu = X_{(sg)}^\mu(s)$ denote a spacelike geodesic parametrized by proper distance $s$. The geodesic begins at event $\mathcal O$ on the central worldline, as shown in Fig.~\ref{fig:timeslices1}.  Thus, $X^\mu_{(cw)}(t_{\mathcal O}) = X_{(sg)}^\mu(0)$. The tangent to the spacelike geodesic is 
\be
    T^\mu = \dot X_{(sg)}^\mu(0)   \ .
\ee
where the dot denotes $d/ds$. 

Expand the coordinates of the point $s$ along the spacelike geodesic  in a series: 
\begin{widetext}
\be
    X_{(sg)}^\mu(s) = X_{(sg)}^\mu(0) + \dot X_{(sg)}^\mu(0) s + \frac{1}{2} \ddot X_{(sg)}^\mu(0) s^2 
    + \frac{1}{6} \dddot X_{(sg)}^\mu(0) s^3 + {\mathcal O}(s^4) \ .
\ee
From the geodesic equation  we have 
\be\label{spacelikegeoeqn}
    \ddot X_{(sg)}^\mu(s) = -\Gamma^\mu_{\alpha\beta}(s) \dot X_{(sg)}^\alpha(s) \dot X_{(sg)}^\beta(s)  \ ,
\ee
where $\Gamma^\mu_{\alpha\beta}(s) \equiv \Gamma^\mu_{\alpha\beta}(X_{(sg)}(s))$ are the 
Christoffel symbols evaluated along the spacelike geodesic. 
By differentiating Eq.~(\ref{spacelikegeoeqn}), we find 
\be
    \dddot X^\mu_{(sg)}(s) = 2\Gamma^\mu_{\alpha\beta}(s) \Gamma^\beta_{\sigma\rho}(s) \dot X^\sigma_{(sg)}(s) \dot X^\rho_{(sg)}(s) \dot X^\alpha_{(sg)}(s) 
    - \partial_\nu \Gamma^\mu_{\alpha\beta}(s) \dot X^\nu_{(sg)}(s) \dot X^\alpha_{(sg)}(s) \dot X^\beta_{(sg)}(s) \ .
\ee
Evaluate these results at $s=0$ to obtain
\bse
\begin{align}
    \ddot X_{(sg)}^\mu(0) &= -\Gamma^\mu_{\alpha\beta}(0) T^\alpha T^\beta \ ,\\
    \dddot X^\mu_{(sg)}(0) &= \left[ 2\Gamma^\mu_{\sigma\rho}(0) \Gamma^\rho_{\alpha\beta}(0) - \partial_\sigma\Gamma^\mu_{\alpha\beta}(0) \right] T^\sigma T^\alpha T^\beta \ .
\end{align}
\ese
Now we have 
\be\label{eqnabove}
    X_{(sg)}^\mu(s) = X_{(sg)}^\mu(0) + T^\mu s - \frac{1}{2} \Gamma^\mu_{\alpha\beta} T^\alpha T^\beta s^2 + 
    \frac{1}{6} \left[ 2\Gamma^\mu_{\sigma\rho} \Gamma^\rho_{\alpha\beta} - \partial_\sigma \Gamma^\mu_{\alpha\beta} \right] T^\sigma T^\alpha T^\beta s^3   + {\mathcal O}(s^4) \ .
\ee
\end{widetext}
The Christoffel symbols (and their derivatives) are evaluated on the 
central worldline at the event ${\mathcal O}$. 

Pick a point on the central worldline, and choose a triad of vectors $e^\mu_{\bar a}$ that is orthogonal to the worldline, and orthonormal. (The index $\bar a$ runs over $1$, $2$, $3$ and labels the legs of the triad.) Extend the triad along the central worldline by Fermi-Walker transport, 
\be
    U^\alpha_{(cw)} \nabla_\alpha e^\mu_{\bar a} = U^\mu_{(cw)} A_\alpha^{(cw)} e^\alpha_{\bar a} \ .
\ee
Expand this out:
\be\label{edotmuaeqn}
    \dot e^\mu_{\bar a} = \dot X_{(cw)}^\mu A_\alpha^{(cw)} e^\alpha_{\bar a} -\Gamma^\mu_{\alpha\beta} \dot X_{(cw)}^\alpha e^\beta_{\bar a} 
\ee
where the dot is $d/d t$. This defines $e^\mu_{\bar a}(t)$, the triad along the central worldline. 
We discretize and evolve this by fourth-order Runge--Kutta. 

Now assume the tangent vector $T^\mu$ is orthogonal to the worldline. We can expand the vector $s T^\mu$ in the triad basis: 
\be
    s T^\mu = x^{\bar a} e^\mu_{\bar a}
\ee
where $x^{\bar a}$ are the expansion coefficients. 

Now reinterpret Eq.~(\ref{eqnabove}) above. The left--hand side is a 
point in spacetime  that depends on the proper time $\bar t$ along the central worldline 
and the coefficients $x^{\bar a}$. We denote this by $x^\mu(\bar t,\bar x)$.  The first term on the right--hand side, $X_{(sg)}^\mu(0)$ is the point ${\mathcal O}$ on the central worldline. In turn, we know the proper time of point ${\mathcal O}$ from the integration of Eq.(~\ref{alphadefappendix}). 
So we can write $X_{(cw)}^\mu(t(\bar t))$ for this term. Likewise, we can write the triad 
vectors in terms of proper time as $e^\mu_{\bar a}(t(\bar t))$. 
Then 
\begin{widetext}
\begin{align}\label{xbartoxtransf}
    x^\mu(\bar t,\bar x) =& X_{(cw)}^\mu(t(\bar t)) + e^\mu_{\bar a}(t(\bar t))  x^{\bar a} 
    - \frac{1}{2} \Gamma^\mu_{\alpha\beta}(t(\bar t)) e^\alpha_{\bar a}(t(\bar t))
    e^\beta_{\bar b}(t(\bar t))  x^{\bar a}  x^{\bar b}  \nono\\
    & + \frac{1}{6} \left[ 2\Gamma^\mu_{\sigma\rho}(t(\bar t)) \Gamma^\rho_{\alpha\beta}(t(\bar t)) - \partial_\sigma\Gamma^\mu_{\alpha\beta}(t(\bar t)) \right] 
    e^\sigma_{\bar c}(t(\bar t)) e^\alpha_{\bar a}(t(\bar t)) e^\beta_{\bar b}(t(\bar t))
    x^{\bar a} x^{\bar b} x^{\bar c}
    + {\mathcal O}(\bar x^4) \ .
\end{align}
This defines a coordinate transformation between generic spacetime coordinates $x^\mu$ and Fermi coordinates $\bar t$, $x^{\bar a}$. 

Now compute derivatives using Eq.~(\ref{edotmuaeqn})
for $\dot e^\mu_{\bar a}$. We have
\bse
\begin{align}
    \frac{\partial x^\mu}{\partial\bar t}  &=  U_{(cw)}^\mu + \left( \delta^\mu_\sigma A_\rho^{(cw)}  - \Gamma^\mu_{\sigma\rho}  \right) U_{(cw)}^\sigma  e^\rho_{\bar a} x^{\bar a} + \left[ \Gamma^\mu_{\alpha\rho} \Gamma^\rho_{\sigma\beta}   - \frac{1}{2} \partial_\sigma \Gamma^\mu_{\alpha\beta} - \Gamma^\mu_{\alpha\sigma} A^{(cw)}_\beta
    \right] U_{(cw)}^\sigma e^\alpha_{\bar a} e^\beta_{\bar b}  x^{\bar a} x^{\bar b} + {\mathcal O}(\bar x^3)   \ ,\\
    \frac{\partial x^\mu}{\partial x^{\bar a}}  &= e^\mu_{\bar a} - \Gamma^\mu_{\alpha\beta} e^\alpha_{\bar a} e^\beta_{\bar b} x^{\bar b}  + \frac{1}{3} \left[ 2\Gamma^\mu_{\sigma\rho}\Gamma^\rho_{\alpha\beta} - \partial_\sigma\Gamma^\mu_{\alpha\beta} + \Gamma^\mu_{\alpha\rho}\Gamma^\rho_{\beta\sigma} - \frac{1}{2}\partial_\alpha\Gamma^\mu_{\beta\sigma}\right] e^\alpha_{\bar a} e^\beta_{\bar b} e^\sigma_{\bar c} x^{\bar b} x^{\bar c} + {\mathcal O}(\bar x^3)   \ ,
\end{align}
\ese
and the inverse relations 
\bse\label{inversecoordtransformation}
\begin{align}
   \frac{\partial \bar t}{\partial x^\mu} =& -U^{(cw)}_\mu 
   + \left[\delta^\beta_\mu A_\alpha^{(cw)} - \Gamma^\beta_{\mu\alpha} \right] U_\beta^{(cw)} e^\alpha_{\bar a} x^{\bar a} \nono\\
   & - \left[ \frac{1}{3} R^\nu{}_{\alpha\beta\sigma}(\delta^\sigma_\mu + U^\sigma_{(cw)} U_\mu^{(cw)} )  + \frac{1}{2} \partial_\mu \Gamma^\nu_{\alpha\beta} - \Gamma^\nu_{\mu\beta} A_\alpha^{(cw)} +  A_\alpha^{(cw)} A_\beta^{(cw)} \delta_\mu^\nu \right] U^{(cw)}_\nu e^\alpha_{\bar a} e^\beta_{\bar b} x^{\bar a} x^{\bar b} 
    + {\mathcal O}(\bar x^3) 
   \ ,\\
   \frac{\partial x^{\bar a}}{\partial x^\mu} =& e^{\bar a}_\mu 
   + \Gamma^\alpha_{\mu\beta} e^{\bar a}_\alpha e^\beta_{\bar b}  x^{\bar b} 
   + \left[ \frac{1}{3} R^\nu{}_{\alpha\beta\sigma} (\delta^\sigma_\mu + U^\sigma_{(cw)} U_\mu^{(cw)}) + \frac{1}{2} \partial_\mu \Gamma^\nu_{\alpha\beta} \right] e^{\bar a}_\nu
   e^\alpha_{\bar b} e^\beta_{\bar c} x^{\bar b} x^{\bar c} + {\mathcal O}(\bar x^3) 
   \ .
\end{align}
\ese
Here, $e_\mu^{\bar a} =  g_{\mu\nu}  \delta^{\bar a \bar b}  e^\nu_{\bar b}$. For notational 
simplicity, we have omitted the arguments $t(\bar t)$ from $\dot X^\mu_{(cw)}$, $U^\mu_{(cw)}$, 
$A^\mu_{(cw)}$, $e^\mu_{\bar a}$. Note that $\Gamma^\mu_{\alpha\beta}$ and $R^\mu{}_{\alpha\beta\nu}$ are also functions of $\bar t$, since they are 
evaluated on the central worldline at the spacetime event   $x^\mu = X^\mu_{(cw)}(t(\bar t))$.

The components of the metric in Fermi coordinates are
\bse
\bea
     g_{\bar t\bar t} (\bar t,\bar x) & = & \frac{\partial x^\mu}{\partial\bar t} g_{\mu\nu}(x(\bar t, \bar x)) \frac{\partial x^\nu}{\partial \bar t} \ , \\
     g_{\bar t \bar a} (\bar t,\bar x) & = & \frac{\partial x^\mu}{\partial\bar t} g_{\mu\nu}(x(\bar t, \bar x)) \frac{\partial x^\nu}{\partial x^{\bar a}} \ , \\
     g_{\bar a\bar b} (\bar t,\bar x) & = & \frac{\partial x^\mu}{\partial x^{\bar a}} g_{\mu\nu}(x(\bar t, \bar x)) \frac{\partial x^\nu}{\partial x^{\bar b}}  \ ,
\eea
\ese
which gives the results listed in Eqs.~(\ref{metricinFermicoords}). 
We also find 
\bse\label{invmetricinFermicoords}
\begin{align}
     g^{\bar t\bar t}(\bar t,\bar x) &= -1 + 2 A_{\bar a}\bigr|_{(cw)} x^{\bar a} 
    + \bigl[ R_{\bar t\bar a\bar t\bar b} -3 A_{\bar a} A_{\bar b} \bigr]\bigr|_{(cw)}  x^{\bar a} x^{\bar b} + {\mathcal O}(\bar x^3) \ ,\\
     g^{\bar t\bar a}(\bar t,\bar x) &= \frac{2}{3} R^{\bar a}{}_{\bar b\bar c\bar t}\bigr|_{(cw)} x^{\bar b} x^{\bar c} + {\mathcal O}(\bar x^3)  \ ,\\
    g^{\bar a\bar b}(\bar t,\bar x) &= \delta^{\bar a\bar b} + \frac{1}{3} R^{\bar a}{}_{\bar c}{}^{\bar b}{}_{\bar d} \bigr|_{(cw)} 
    x^{\bar c} x^{\bar d} +  {\mathcal O}(\bar x^3)  \ .
\end{align}
\ese
for the inverse metric. Here we are using notation such as $A_{\bar a}\bigr|_{(cw)} \equiv A^{(cw)}_\alpha e^\alpha_{\bar a}$ and  $R_{\bar t\bar a\bar t\bar b}\bigr|_{(cw)} \equiv  R_{\mu\alpha\nu\beta} U^{\mu}_{(cw)} e^\alpha_{\bar a} U^\nu_{(cw)} e^\beta_{\bar b}$ (with $R_{\mu\alpha\nu\beta}$  evaluated at $x^\mu = X^\mu_{(cw)}(t(\bar t))$) to denote the 
Fermi components of the acceleration and Riemann tensor on the central worldline. 

To compute the conserved energy and angular momentum, we need the Fermi coordinate components of a covector field. For $V_\mu(x)$, we find  
\bse\label{covvecinFermicoords}
\begin{align}
    V_{\bar t}(\bar t,\bar x) = &  V_{\bar t} \bigr|_{(cw)} 
    + \bigl[\nabla_{\bar a} V_{\bar t} + A_{\bar a}  V_{\bar t}\bigr]\bigr|_{(cw)}    x^{\bar a} \nono\\
    & +\frac{1}{2}  \bigl[ \nabla_{\bar a}\nabla_{\bar b} V_{\bar t} 
    + R^{\bar t}{}_{\bar a\bar b\bar t} V_{\bar t} 
    + R^{\bar c}{}_{\bar a\bar b\bar t} V_{\bar c} + 2A_{\bar a} \nabla_{\bar b} V_{\bar t} \bigr]\bigr|_{(cw)}
     x^{\bar a} x^{\bar b}  + {\mathcal O}(\bar x^3) \ ,\\
    V_{\bar a}(\bar t,\bar x) = & V_{\bar a}\bigr|_{(cw)} 
    +  \nabla_{\bar b} V_{\bar a} \bigr|_{(cw)} x^{\bar b} + \frac{1}{6}\bigl[ 3 \nabla_{\bar b} \nabla_{\bar c} V_{\bar a} + R^{\bar t}{}_{\bar b\bar c\bar a} V_{\bar t} 
     + R^{\bar d}{}_{\bar b\bar c\bar a} V_{\bar d}\bigr] \bigr|_{(cw)} x^{\bar b} x^{\bar c}  + {\mathcal O}(\bar x^3)  \ ,
\end{align}
\ese
where $V_{\bar t}\bigr|_{(cw)} \equiv V_\mu U^\mu_{(cw)}$ and $V_{\bar a} \bigr|_{(cw)} \equiv V_\mu e^\mu_{\bar a} $ 
(with $V_\mu$ evaluated at $x^\mu = X^\mu_{(cw)}(t(\bar t))$).

\end{widetext} 
\bibliography{ref.bib}
\bibliographystyle{unsrt}
\end{document}